\documentclass[letterpaper,10pt]{article}

\usepackage[utf8]{inputenc}
\usepackage{amsmath}
\usepackage{amssymb}
\usepackage{amsfonts}
\usepackage{psfrag}
\usepackage{graphicx}
\usepackage{epsfig}

\usepackage[stable]{footmisc}

\newcommand{\dirac}{{\slash \negthinspace \negthinspace \negthinspace \nabla}}

\title{Quasinormal frequencies of a two-dimensional asymptotically anti-de Sitter black hole of the dilaton gravity theory}

\author{M.\ I.\ Hern\'andez-Vel\'azquez$^{1}$, A.\ L\'opez-Ortega$^{1,*}$ \\
$^{1}$Departamento de F\'{\i}sica, 
Escuela Superior de F\'{\i}sica y Matem\'aticas, \\
Instituto Polit\'ecnico Nacional, 
Ciudad de M\'exico, M\'exico. \\
Correspondence$^{*}$: A.\ L\'opez-Ortega\\
alopezo@ipn.mx
}

\begin{document}

\maketitle

\begin{abstract}

We numerically calculate the quasinormal frequencies of the Klein-Gordon and Dirac fields propagating in a two-dimensional asymptotically anti-de Sitter  black hole of the dilaton gravity theory. For the Klein-Gordon field we use the Horowitz-Hubeny method and the asymptotic iteration method for second order differential equations. For the Dirac field we first exploit the Horowitz-Hubeny method. As a second method, instead of using the asymptotic iteration method for second order differential equations, we propose to take as a basis its formulation for coupled systems of first order differential equations. For the two fields we find that the results that produce the two numerical methods are consistent. Furthermore for both fields we obtain that their quasinormal modes  are stable and we compare their quasinormal frequencies to analyze whether their spectra are isospectral. Finally we discuss the main results.


\end{abstract}

\section{Introduction}
\label{s: Introduction}

The quasinormal modes (QNMs) are the oscillations of perturbed black holes that are purely ingoing near the horizon. The boundary condition imposed at the asymptotic region depends on its structure, for example, for asymptotically flat black holes the boundary condition usually imposed is that the perturbation is purely outgoing as $r \to \infty$, whereas for the asymptotic anti-de Sitter black holes (asymptotically adS black holes, in what follows), a common boundary condition  demands that the perturbation goes to zero  as $r \to \infty$ \cite{Kokkotas:1999bd}--\cite{Avis:1977yn}. It is convenient to notice that in asymptotically adS spacetimes we can impose a different boundary condition as $r \to \infty$. See Refs.\ \cite{Breitenlohner:1982bm}--\cite{Dasgupta:1998jg} for some examples. Associated with the QNMs we find a set of complex frequencies called quasinormal frequencies (QNFs). It is well known that the QNFs of black holes are determined by the geometry and the type of perturbation \cite{Kokkotas:1999bd}--\cite{Berti:2009kk}. 

Two dimensional gravity allows us to analyze several physical problems in a simple framework. The physical properties of two-dimensional  black holes (2D black holes, in what follows) are studied because some aspects of the analysis are simpler than in spacetimes with $D \geq 3$ dimensions \cite{Grumiller:2002nm}, \cite{Grumiller:2006rc}. Furthermore, in 2D spacetimes the equations of motion for classical fields simplify and we can study in more detail several physical phenomena, for example, the way in which a 2D black hole reacts when it is perturbed. The QNFs of 2D spacetimes have been studied in recent times \cite{Becar:2007hu}--\cite{Zelnikov:2008rg}.  Exact results for the QNFs have previously found \cite{Becar:2007hu}--\cite{Zelnikov:2008rg}, but for some asymptotically adS 2D backgrounds a numerical calculation is necessary \cite{Cordero:2012je}. 

We notice that the QNFs of asymptotically adS black holes have attracted attention, since they are useful to study its classical stability \cite{Kokkotas:1999bd}--\cite{Berti:2009kk}, \cite{Cordero:2012je}, \cite{Horowitz:1999jd}--\cite{Giammatteo:2004wp} and to calculate the entropy spectrum of the black hole horizon \cite{Hod:1998vk}--\cite{LopezOrtega:2010wx}. Here we exploit the advantages that we find in 2D gravity to implement two numerical methods to  calculate the QNFs of a 2D asymptotically adS black hole \cite{Lemos:1993py}. Additionally, we carry out this work  to understand the properties of this 2D black hole of Ref.\ \cite{Lemos:1993py}  and study its classical stability under small perturbations.  One of the numerical procedures is the widely used Horowitz-Hubeny method (HH method in what follows) \cite{Horowitz:1999jd} that is appropriate to compute the QNFs of asymptotically adS spacetimes \cite{Konoplya:2011qq}, \cite{Berti:2009kk}. The other numerical procedure is the asymptotic iteration method (AIM in what follows) \cite{Ciftci Hall and Saad 2003}, that is modified to determine the QNFs of black holes  \cite{Cho:2009cj}. For the Klein-Gordon field the AIM is exploited in the usual formulation for second order differential equations \cite{Ciftci Hall and Saad 2003}, \cite{Cho:2009cj}, whereas for the Dirac field we propose to use the formulation of the AIM for coupled  systems of first order differential equations \cite{Cifti:2005 coupled}. As far as we know, this version of the AIM has not previously used to calculate QNFs of black holes and we think that the computation of the QNFs for 2D black holes is the appropriate setting to test this method. 

The rest of the paper is structured as follows. In Sect.\ \ref{ss: black hole 2D} we give the asymptotically adS 2D black hole that we study in this work and we describe its main properties. In Sect.\ \ref{s: Klein-Gordon qnms} we simplify the equation of motion for the Klein-Gordon field in the 2D black hole that we study. Furthermore, for the Klein-Gordon field  in Sects.\ \ref{ss: HH Klein-Gordon} and \ref{ss: aim Klein-Gordon} we describe the HH method and the AIM for second order differential equations. For the Dirac field we simplify its equation of motion in Sect.\ \ref{s: Dirac qnms}. Also, for the Dirac field our implementations of   the HH method and the AIM for coupled systems of first order differential equations are outlined in Sects.\ \ref{ss: HH Dirac} and \ref{ss: aim Dirac}. In Sect.\ \ref{ss: numerical Klein-Gordon} we numerically calculate the QNFs of the Klein-Gordon field moving in the asymptotically adS 2D black hole that we are studying and then the QNFs of the Dirac field are numerically computed in Sect.\  \ref{ss: numerical Dirac}.  For the two fields we compare the results that produce the two methods. In Sect.\ \ref{s: Discussion} we discuss the main results. For the Klein-Gordon and Dirac fields, in Appendix \ref{a: additional tables} we give the numerical values of the QNFs with more decimal places. For static 2D spacetimes with diagonal metric in Appendix \ref{a: null dyad}  we simplify the Dirac equation to a pair of Schr\"odinger type equations. The simplification uses a basis of null vectors. This method can be exploited in other 2D spacetimes, and as far as we know, it is not already reported. In Appendix \ref{a: KG coupled AIM} we extend the discussion of the AIM for the Klein-Gordon field and  we show how to use the formulation of the AIM for coupled systems of first order differential equations  to determine its QNFs. In Appendix \ref{a: iaim coupled} we give the so-called improved formulation of the AIM for second order differential equations \cite{Cho:2009cj}, \cite{Cho:2011sf} and we employ this improved version to calculate the QNFs of the Klein-Gordon field. Furthermore, we develop an improved version of the AIM  for coupled systems of first order differential equations. Taking as a basis this improved version we calculate the QNFs of the Dirac field again and we compare with the previous results. Finally in Appendix \ref{a: Lifshitz qnfs} we show that the  AIM  for coupled systems of first order differential equations works for calculating the QNFs of the Dirac field moving in a higher dimensional Lifshitz black hole.

\section{Methods}

In this section we describe the main properties of the asymptotically adS 2D black hole that we study in this work. Furthermore we give the  steps to simplify the equations of motion for the Klein-Gordon field and the Dirac field in the black hole under study. Finally, for both fields, we summarize the HH method and the AIM.

\subsection{Two-dimensional asymptotically anti-de Sitter black hole}
\label{ss: black hole 2D}

In Ref.\ \cite{Lemos:1993py} we find several 2D spacetimes representing  black holes. These black holes are solutions to the equations of motion for the action 
\begin{equation} \label{e: action}
S = \frac{1}{2\pi} \int{d^{2}x\sqrt{-g}e^{-2\phi}[\mathcal{R}-4 \rho (\partial\phi)^2 + 4 \sigma^2]} ,
\end{equation} 
where $\mathcal{R}$ is the scalar curvature, $\sigma$ is a constant that we interpret as a cosmological constant, $\phi$ is the dilaton, and $\rho$ is a parameter that we take as $\rho = -1/2$. For this value of $\rho$, a solution to the equations of motion is the  asymptotically adS 2D black hole that we study in this work and whose metric is \cite{Lemos:1993py} 
\begin{equation} \label{e: black hole two-dimensional}
d s^2 = \left(a^2 r^2 - \frac{1}{ar}\right) d t^2 - \left(a^2 r^2 - \frac{1}{ar}\right)^{-1} d r^2 ,
\end{equation} 
and the dilaton field is equal to
\begin{equation}
 e^{-2\phi} = a^2r^2 .
\end{equation} 
The event horizon of this black hole is located at $r_+ = 1/a$, where $a$ is related to $\sigma$ by $a=\sqrt{2} |\sigma|/\sqrt{3}$. We notice that the scalar curvature of this 2D black hole is equal to $\mathcal{R}=-4 \sigma^2/3 + 4 \sigma^2/(3 a^3 r^3)$, hence it is positive inside the horizon and negative outside the horizon \cite{Lemos:1993py}. The ADM mass of the 2D black hole (\ref{e: black hole two-dimensional}) is equal to $M=|\sigma|/\sqrt{6}$ \cite{Lemos:1993py}, \cite{Lemos:1994fn}.  We observe that the metric of the 2D  black hole (\ref{e: black hole two-dimensional}) is the $(t,r)$ sector of the metric for the flat horizon Schwarzschild adS black hole in four dimensions \cite{Lemos:1994fn}, but we point out that the behavior of the scalar curvature is different in both solutions.  Owing this fact we choose to analyze the stability under perturbations of this asymptotically adS 2D black hole.

As far as we know, the classical stability under perturbations of this 2D black hole has not previously analyzed and to begin this study we calculate its spectrum of QNFs. We have not been able to find exact solutions to the equations of motion for the Klein-Gordon and Dirac fields, hence, in what follows we use the HH method \cite{Horowitz:1999jd} and the AIM \cite{Ciftci Hall and Saad 2003}, \cite{Cho:2009cj}, \cite{Cifti:2005 coupled} to compute numerically the frequencies of its damped oscillations. Furthermore, we evaluate whether the AIM for coupled systems of first order differential equations works to produce the QNFs of the Dirac field.

As usual for asymptotically adS black holes,  we define the QNMs as the oscillations that satisfy the boundary conditions \cite{Kokkotas:1999bd}--\cite{Berti:2009kk}, \cite{Cordero:2012je}, \cite{Horowitz:1999jd}--\cite{Giammatteo:2004wp}\\
a) The  field is purely ingoing near the horizon.\\
b) The field goes to zero as $r \to \infty$.\\
Note that in the boundary condition b) we assume that the coordinates of the line element (\ref{e: black hole two-dimensional}) are used.

\subsection{Equation of motion for the Klein-Gordon field}
\label{s: Klein-Gordon qnms}

In the 2D black hole (\ref{e: black hole two-dimensional}) we first calculate the QNFs of a test Klein-Gordon field whose equation of motion is
\begin{equation} \label{e: Klein-Gordon}
(\square + m ^2)\Phi = 0.
\end{equation}
As shown in Ref.\  \cite{LopezOrtega:2011sc}, if we take a separable solution 
\begin{equation} \label{e: ansatz Klein-Gordon}
\Phi = R(r) e ^{-i\omega t},
\end{equation}
then the Klein-Gordon equation simplifies to  a Schr\"odinger type equation 
\begin{equation} \label{e: Schrodinger type Klein-Gordon}
\frac{d^2 R}{d r_{*}^2} +\omega^2 R =V_{KG} R,
\end{equation}
with $r_*$ denoting the tortoise coordinate of the 2D black hole (\ref{e: black hole two-dimensional})
\begin{equation}
 r_* = \int d r \left(a^2 r^2 - \frac{1}{ar}\right)^{-1}  ,
\end{equation} 
and the effective potential $V_{KG}$ is equal to \cite{LopezOrtega:2011sc}
\begin{equation} \label{e: potential Klein Gordon} 
V_{KG}=m^2 \left(a^2 r^2-\frac{1}{ar}\right).
\end{equation}
From this expression for the effective potential $V_{KG}$, we note that for $m=0$, it goes to zero. Therefore for $m=0$ the solutions of Eq.\ (\ref{e: Schrodinger type Klein-Gordon}) are sinusoidal functions and since we are interested in damped solutions, in what follows we consider a test massive Klein-Gordon field with $m > 0$. Notice that for our problem the physically relevant interval is $r \in (r_+,\infty)$. It is convenient to comment that in the 2D black hole (\ref{e: black hole two-dimensional}) we have not been able to solve exactly the radial equation (\ref{e: Schrodinger type Klein-Gordon}) and therefore we numerically compute the QNFs of the Klein-Gordon field. Furthermore, we assume the test field approximation for calculating the QNFs of the asymptotically adS 2D black hole (\ref{e: black hole two-dimensional}).

We point out that near the event horizon at $r_+ = 1/a$ the effective potential $V_{KG}$ goes to zero and therefore the radial function near the horizon  behaves as 
\begin{equation} \label{e: near horizon Klein}
 R \approx K_{1} \,\,\textrm{e}^{+i \omega r_{*}} + K_{2}\,\,\textrm{e}^{-i \omega r_{*}},
\end{equation}
where $K_{1}$, $K_{2}$ are constants. Note that near the horizon the first term of the previous formula is an outgoing wave and the second term is an ingoing wave.

To satisfy  the QNMs boundary condition near the horizon, we impose that the field is purely ingoing near the horizon by taking the radial function $R$ as \cite{Horowitz:1999jd}
\begin{equation} \label{e: ansatz R Klein}
R(r)= e^{-i\omega{r}_{*} } \tilde{R} (r),                     
\end{equation}
to get that the function $\tilde{R}$ must be a solution of the differential equation 
\begin{equation} \label{e: radial ingoing Klein}
f \frac{{d}^{2} \tilde{R}}{{d r}^2 }+\left(\frac{d f}{d r}-2i\omega\right) \frac{d \tilde{R}}{d r}-\frac{V}{f} \tilde{R}=0 ,
\end{equation}
where we define  $f = a^2 r^2 - 1/ (a r) $.

\subsection{Horowitz-Hubeny method for the Klein-Gordon field}
\label{ss: HH Klein-Gordon}

The numerical HH method proposed in \cite{Horowitz:1999jd} is widely used to compute the QNFs of asymptotically adS black holes \cite{Konoplya:2011qq}, \cite{Berti:2009kk}, \cite{Cordero:2012je}. Following Horowitz-Hubeny, to transform the interval $(r_+,\infty)$ into a finite interval, we make the change of variable 
\begin{equation} \label{e: change x r}
 x=\frac{1}{r},
\end{equation} 
to obtain that Eq.\  (\ref{e: radial ingoing Klein}) transforms into a differential equation of the form \cite{Horowitz:1999jd}
\begin{equation} \label{e: equation s t u Klein}
   \tilde{s}(x) \frac{{d}^{2} \tilde{R}}{d {x}^2 }+\frac{\tilde{t}(x)}{x-{x}_{+}}\frac{d \tilde{R}}{d x}+\frac{u(x)}{{(x-{x}_{+})}^2} \tilde{R}=0,
\end{equation}
where
\begin{equation} \label{e: inverse radius}
x_{+}= \frac{1}{r_{+}} = a ,
\end{equation}
and the functions $\tilde{s}$, $\tilde{t}$, and  $u$ are equal to
\begin{eqnarray} \label{e: s t u Klein}
\tilde{s}(x)&=&{x}^{2} ({a}^{2}+ax+{x}^{2}), \nonumber \\
\tilde{t}(x)&=&\left(\frac{3x^{2}}{a}-2i\omega\right){x}^{2} a, \\
u(x)&=& a {m}^{2}(x-a). \nonumber 
\end{eqnarray}
Notice that the function $u$ satisfies $u(a) = 0$.

To preserve  the  purely ingoing radial solution near the horizon,  we expand the function $\tilde{R}$ in the form \cite{Horowitz:1999jd}
\begin{equation} \label{e: radial expansion horizon Klein}
\tilde{R} (x)= \sum_{k=0}^{\infty}{a}_{k}(\omega){(x-{x}_{+})}^{k}.
\end{equation}
Furthermore, we expand the function $\tilde{s}$ as
\begin{equation}
 \tilde{s}(x)=\sum_{k=0}^{\infty}\tilde{s}_{k}(\omega){(x-{x}_{+})}^{k} ,
\end{equation}
and we make analogous expansions for the functions $\tilde{t}$ and $u$.

Substituting these expansions into the differential equation (\ref{e: equation s t u Klein}) we obtain that the coefficients ${a}_{k}(\omega)$ are given by the recurrence relation \cite{Horowitz:1999jd}
\begin{equation} \label{e: coefficients a}
{a}_{k}= -\frac{1}{k(k-1)\tilde{s}_{0}+k\tilde{t}_{0}} \sum_{n=0}^{k-1} {a}_{n}( n(n-1)\tilde{s}_{k-n}+n\tilde{t}_{k-n}+u_{k-n}).
\end{equation}
Since $a_0$ is not determined by the method, in what follows we take $a_0 = 1$ \cite{Horowitz:1999jd}.

As $r \to \infty $ the QNMs boundary condition b) imposes that the radial function goes to zero. From the expansion (\ref{e: radial expansion horizon Klein}), to fulfill this boundary condition and to find the QNFs of the field we must solve 
\begin{equation} \label{e: condition infinity QNM}
\sum_{k=0}^{\infty}{a}_{k}(\omega){(-{x}_{+})}^{k}=0.
\end{equation}
We see that the roots of this equation are the QNFs of the 2D black hole \eqref{e: black hole two-dimensional}. Note that we cannot compute the infinite sum of the previous formula, therefore a common method to solve Eq.\  (\ref{e: condition infinity QNM}) is to calculate the sum up to an integer value $N$ and obtain the roots of the resulting polynomial. We repeat the calculation for another integer value $N_1 > N$ and the common roots are the QNFs of the Klein-Gordon field \cite{Horowitz:1999jd}. In the HH method the repeated roots for different values of $N$ are called stable roots.

\subsection{Asymptotic iteration method for the Klein-Gordon field}
\label{ss: aim Klein-Gordon}

The asymptotic iteration method  is useful to study linear second order differential equations that can be written in the form \cite{Ciftci Hall and Saad 2003}
\begin{equation}\label{eq:aim}
\phi^{\prime \prime} = \lambda_0 \phi^{\prime}+s_0\phi,
\end{equation}
where $\lambda_0$ and $s_0$ are differentiable functions of the independent variable $x$. As usual, we denote the derivative with respect to the independent variable $x$ with a prime. This method is widely used to solve linear second order differential equations, to find its eigenvalues  \cite{Ciftci Hall and Saad 2003},  and more recently is taken as a basis to calculate the QNFs of black holes \cite{Cho:2009cj}.  

We observe that the derivative of the previous equation takes the form \cite{Ciftci Hall and Saad 2003}
\begin{equation}\label{eq: first derivative}
\phi^{\prime \prime \prime} = \lambda_1 \phi^{\prime} +s_1\phi,
\end{equation}
where 
\begin{equation}\label{eq: recurrence 1}
\lambda_1 =\lambda_0^{\prime} + s_0 + \lambda_0^2, \quad  \quad s_1=s_0^{\prime} + s_0 \lambda_0 ,
\end{equation}
that is, for the differential equation (\ref{eq:aim}), the structure of its first derivative is similar when we define $\lambda_1$ and $s_1$ as previously \cite{Ciftci Hall and Saad 2003}. Furthermore, we find that the $n$-th derivative of Eq.\ (\ref{eq:aim}) takes an analogous form \cite{Ciftci Hall and Saad 2003}
\begin{equation}\label{eq:nth derivative}
\phi^{(n+2)}= \lambda_{n} \phi^{\prime}+s_{n} \phi,
\end{equation} 
where 
\begin{equation}\label{eq: recurrence general}
\lambda_n =\lambda_{n-1}^{\prime} + s_{n-1} + \lambda_0 \lambda_{n-1}, \quad  \quad s_n=s_{n-1}^{\prime} +s_0 \lambda_{n-1}.
\end{equation}
 
To find a solution to Eq.\ (\ref{eq:aim}), in Ref.\ \cite{Ciftci Hall and Saad 2003} is imposed the asymptotic aspect of the AIM by proposing that for $n$ sufficiently large the following equation \cite{Ciftci Hall and Saad 2003}
\begin{equation}\label{eq:condition 1}
\frac{s_n}{\lambda_n} =\frac{s_{n-1}}{\lambda_{n-1}} = \alpha,
\end{equation}
is satisfied. From the previous equation we obtain the \textit{quantization condition}\footnote{In this work, for Eq.\ (\ref{eq: quantization}) we use the name of quantization condition \cite{Ciftci Hall and Saad 2003}, but notice that here we explore the classical propagation of the Klein-Gordon and Dirac fields. In the context of this paper a more appropriate name for Eq.\ (\ref{eq: quantization}) could be discretization condition or termination condition.} \cite{Ciftci Hall and Saad 2003}
\begin{equation}\label{eq: quantization}
\delta_n= \lambda_n s_{n-1} - \lambda_{n-1} s_n = 0. 
\end{equation}
Usually the previous condition depends on the independent variable $x$ and the oscillation frequency. To get the QNFs from this condition we evaluate the quantity $\delta_n$ in a convenient point and the stable roots of the resulting equation are the QNFs \cite{Ciftci Hall and Saad 2003}. In the AIM for stable roots we understand the common roots of the quantization condition (\ref{eq: quantization}) for different values of $N$ where $N$ is the number of times we iterate the expressions (\ref{eq: recurrence general}). Usually to determine the stable roots we take values of $N$ that differ by five, that is, we calculate the quantity $\delta_N$ of Eq.\ (\ref{eq: quantization}) for two values of $N$ that differ by 5 and we take the common roots of Eqs.\ (\ref{eq: quantization}) as the QNFs.  We point out that to implement the AIM we scale out the behavior of the field near the boundaries before we find the equivalent of Eq.\ (\ref{eq:aim}) for our problem \cite{Cho:2009cj}.

In the following we use the AIM to calculate the QNFs of the Klein-Gordon field propagating in the 2D black hole (\ref{e: black hole two-dimensional}). We take as a basis Eq.\ (\ref{e: radial ingoing Klein}) and we factor out the behavior at the asymptotic region, since the behavior near the horizon is scaled out. As a first step we study Eq.\ (\ref{e: radial ingoing Klein}) as $r \to \infty$ to get
\begin{equation}
 \frac{d^2 \tilde{R} }{dr^2} + \frac{2}{r} \frac{d \tilde{R}}{d r} -\left(\frac{m^2}{a^2}  \right) \frac{\tilde{R}}{r^2} = 0 ,
\end{equation} 
whose solutions are of the form
\begin{equation} \label{e: asymptotic KG}
 \tilde{R}  =  K_{1}   r^{q_-} + K_{2}  r^{q_+} ,
\end{equation} 
where $K_{1}$, $K_{2}$ are constants and  the quantities $q_-$, $q_+$ are equal to
\begin{equation}
q_- = -1/2 - \sqrt{1 + 4 m^2 r_+^2}/2, \qquad \qquad  q_+ = -1/2 + \sqrt{1 + 4 m^2 r_+^2}/2.
\end{equation} 
From the expression (\ref{e: asymptotic KG}) we observe that the solution satisfying the boundary condition of the QNFs as $r \to \infty$ is  $K_1 r^{q_-}$. Therefore we make the ansatz 
\begin{equation}
 \tilde{R} = r^{q_-} \hat{R}
\end{equation} 
to get that the function $\hat{R}$ satisfies the differential equation
\begin{align} \label{e:  radial infinity KG}
 f \frac{d^2 \hat{R} }{d r^2} &+ \left( \frac{2 q_- f }{r} +  \frac{d f }{d r} - 2 i \omega \right)  \frac{d \hat{R} }{d r} \nonumber \\
  &+ \left( \frac{ q_- (q_- -1) f}{r^2} + \frac{q_-}{r} \frac{d f }{d r} - \frac{2 i \omega q_-}{r} - \frac{V}{f} \right) \hat{R} = 0 .
\end{align}

Owing to the radial variable is defined in a semi-infinite interval, in a similar way to the HH method, it is convenient to use a new independent variable with a finite range. For implementing the AIM we find useful to define the variable
\begin{equation} \label{e: z variable}
 z=\frac{r_+}{r},
\end{equation} 
such that $z=1$ for $r=r_+$ and $z= 0$ as $r \to \infty$. Taking as independent variable to $z$, we obtain that Eq.\ (\ref{e:  radial infinity KG}) transforms into
\begin{align} \label{e: radial AIM Klein-Gordon}
 &\frac{d^2 \hat{R} }{d z^2}- \left(  \frac{2 q_-}{z} + \frac{2+z^3}{z(1-z^3)} - \frac{2}{z} - \frac{2 i \omega r_+}{1-z^3} \right)\frac{d \hat{R} }{d z}  \\
 &-\left( \frac{m^2 r_+^2}{z^2 (1-z^3)}- \frac{q_- (q_--1)}{z^2} -\left( \frac{2+z^3}{z}- 2 i \omega r_+ \right)\frac{q_-}{z(1-z^3)} \right)\hat{R} =0 . \nonumber
\end{align}
The previous linear second order differential equation is of the form (\ref{eq:aim}) and for the Klein-Gordon field we take this equation as a basis to implement the AIM.  Thus, we identify the functions $\lambda_0$, $s_0$ as follows
\begin{align} \label{e: lambda 0 ese 0}
 \lambda_0 (z) &= \frac{2 q_-}{z} + \frac{2+z^3}{z(1-z^3)} - \frac{2}{z} - \frac{2 i \omega r_+}{1-z^3},  \\
 s_0 (z) &= \frac{m^2 r_+^2}{z^2 (1-z^3)}- \frac{q_- (q_--1)}{z^2} -\left( \frac{2+z^3}{z}- 2 i \omega r_+ \right)\frac{q_-}{z(1-z^3)}, \nonumber
\end{align}
and we use the recurrence relations (\ref{eq: recurrence general}) to calculate the quantities that appear in the quantization condition (\ref{eq: quantization}). The stable roots of the condition (\ref{eq: quantization}) are the QNFs of the Klein-Gordon field. We note that in Sect.\ \ref{ss: numerical Klein-Gordon} to calculate the QNFs of the Klein-Gordon field we follow the procedure described in Ref.\ \cite{Ciftci Hall and Saad 2003} instead of the improved AIM proposed in Ref.\ \cite{Cho:2009cj}, (but see Appendix \ref{a: iaim coupled}).

\subsection{Equation of motion for the Dirac field}
\label{s: Dirac qnms}

The equation of motion for a Dirac field is
\begin{equation} \label{e: Dirac equation}
 i \dirac \Psi = m \Psi ,
\end{equation} 
where $\dirac$ is the Dirac operator and $m$ is the mass of the field. In what follows we assume that $m > 0$. Reference \cite{LopezOrtega:2011sc} shows that in a static  2D spacetime whose metric is diagonal, the Dirac equation in the chiral representation simplifies to a pair of Schr\"odinger type equations with effective potentials 
\begin{equation} \label{e: potentials Dirac square root}
 V_\pm = m^2 f \mp \frac{m}{2} \sqrt{f} \frac{d f}{d r} .
\end{equation} 
Owing to the factor $\sqrt{f}$ in these effective potentials, our numerical code for the HH method converges slowly.\footnote{We note that the effective potentials (\ref{e: potentials Dirac square root}) are supersymmetric partners. The superpotential is equal to $W = - m \sqrt{f}$ \cite{LopezOrtega:2011sc}, \cite{Cooper:1994eh}.} Thus, to get a different pair of effective potentials for the radial equations of the Dirac field, in what follows, to write the Dirac equation in the 2D black hole (\ref{e: black hole two-dimensional}), we exploit a dyad of null vectors (see Appendix \ref{a: null dyad} for more details). 

In this basis of null vectors, the Dirac equation simplifies to a pair of Schr\"odinger type equations 
\begin{equation} \label{e: Schrodinger type Dirac}
\frac{d^{2} R_{l}}{d r_{*}^{2}} + \omega^2 R_{l} = V_{l}R_{l},
\end{equation}
with $l=1,2$, and the effective potentials are equal to  
\begin{equation} \label{e: Potentials Dirac 1/2}
V_{l} = m^2 f \mp \frac{i \omega}{2}\frac{d f}{d r} - \frac{f}{4}\frac{d ^{2} f}{d r^{2}} + \frac{1}{16}\left(\frac{d f}{d r}\right)^{2} .
\end{equation}
In the previous equation and in what follows, the upper (lower) sign corresponds to $l=1$ ($l=2$). We notice that the two effective potentials (\ref{e: Potentials Dirac 1/2}) do not contain square roots of the function $f$. Also the system of coupled equations (\ref{e: coupled radial odes}) for the components of the spinor is an appropriate basis to use the AIM for coupled systems (see Sect.\ \ref{ss: aim Dirac}).

Near the horizon of the black hole, the effective potentials (\ref{e: Potentials Dirac 1/2}) simplify to 
\begin{equation} \label{e: near horizon potential Dirac}
\lim_{r \to r_{+}} V_{l} = \mp \frac{i\omega}{2}\left. \frac{d f}{d r} \right|_{r = r_{+}} + \frac{1}{16}\left( \left.  \ \frac{d f}{d r}\right|_{r = r_{+}} \right)^{2},
\end{equation}
and therefore near the horizon the radial functions $R_{l}$ behave in the form 
\begin{equation} \label{e: near horizon Dirac}
R_{l} \approx K^{I}_{l} \,\,\textrm{e}^{ {-i}\left( \omega  \pm \frac{i \kappa}{2} \right) r_{*} } 
+K^{II}_{l} \,\,\textrm{e}^{ +{i}\left( \omega  \pm \frac{i \kappa }{2} \right)r_{*} },
\end{equation}
where $K^{I}_{l}$, $K^{II}_{l}$ are constants and in the previous equation we introduce the surface gravity of the 2D black hole (\ref{e: black hole two-dimensional})  defined by
\begin{equation} \label{e: surface gravity}
\kappa = \frac{1}{2} \left. \frac{d f}{d r} \right|_{r = r_{+}} = \frac{3a}{2}.
\end{equation}
Also, we note that near the horizon the first term of the expression (\ref{e: near horizon Dirac}) is an ingoing wave, whereas the second term is an outgoing wave.

We have not been able to solve exactly the Schr\"odinger type equations (\ref{e: Schrodinger type Dirac}) and therefore we use first the HH method to compute the QNFs of the Dirac field in the 2D black hole (\ref{e: black hole two-dimensional}). Thus, in the following subsection, taking as a basis the Schr\"odinger type equations (\ref{e: Schrodinger type Dirac}) with the effective potentials (\ref{e: Potentials Dirac 1/2}) and the HH method, we calculate the QNFs of the Dirac field in the 2D black hole (\ref{e: black hole two-dimensional}). Moreover, in Sect.\ \ref{ss: aim Dirac} we show how to use the AIM for coupled systems of first order equations to compute the QNFs of the Dirac field.

\subsection{Horowitz-Hubeny method for the Dirac field}
\label{ss: HH Dirac}

To use the HH method \cite{Horowitz:1999jd} for calculating the QNFs of the Dirac field we make the following transformations to the differential equations (\ref{e: Schrodinger type Dirac}). Near the horizon, to satisfy  the boundary condition  of the QNMs we  take the functions $R_{l}$ in the form 
\begin{equation} \label{e: first change radial Dirac}
R_{l} = e^{-i\left( \omega \pm \frac{i\kappa}{2}\right) r_{*}} \tilde{R}_{l}(r),
\end{equation} 
to get that the functions $\tilde{R}_{l}$ must be solutions of the differential equations
\begin{equation} \label{e: Dirac radial first}
f \frac{{d}^{2} \tilde{R}_{l}}{{d r}^2 }+\left(\frac{d f}{d r}-2i\omega \pm  \kappa  \right) \frac{d \tilde{R}_{l}}{d r}+\left(\frac{\kappa^{2}}{4} \mp i\kappa\omega -V_{l}  \right)\frac{1}{f} \tilde{R}_{l}=0 ,
\end{equation}
that transform into 
\begin{align} \label{e: Dirac radial x}
x^{4}f \frac{d^{2} \tilde{R}_{l}}{d x^{2}} & +\left(x^{4}\frac{d f}{d x} + 2x^{3}f + 2i\omega x^{2}   \mp  \kappa x^{2}\right)  \frac{d \tilde{R}_{l}}{d x}   \nonumber \\ 
& + \left(\frac{\kappa^{2}}{4} \mp i\kappa\omega -V_{l}  \right) \frac{1}{f} \tilde{R}_{l} = 0,
\end{align}
when we use the variable $x$ defined in the formula (\ref{e: change x r}).

In a straightforward way we transform Eqs.\  (\ref{e: Dirac radial x}) to the form (\ref{e: equation s t u Klein}), but the functions $\tilde{s}_{l}$, $\tilde{t}_{l}$, and $u_{l}$ for the Dirac field are given by
\begin{align}
\tilde{s}_{l} (x) &= 16x^{2}(x^{2}+ax+a^{2})^{2} ,\nonumber \\
\tilde{t}_{l} (x) &=16a (x^{2}+ax+a^{2}) \left(\frac{3x^{4}}{a} -2i\omega x^{2} \pm  \kappa x^{2}\right) , \\
u_{l} (x)&= 9a^{4}x^{2} \mp 24i\omega a^3 x^{2} + 8a(2m^{2} -3a^{2})(x^3 - a^3) \nonumber \\ 
& + 8\left( x^{3} + 2a^{3} \right)  \left( x^3 - a^{3} \pm i \omega ax  \right)  - \left( x^{3} + 2a^{3}  \right)^{2}. \nonumber
\end{align}
We notice that $u_{l}(a) = 0$. From these expressions and taking into account the recurrence relation (\ref{e: coefficients a}) we get the coefficients $a_k$ for the Dirac field moving in the 2D black hole (\ref{e: black hole two-dimensional}). As previously, the QNFs of the Dirac field are the stable roots of the corresponding equations that have the same mathematical form as Eq.\ (\ref{e: condition infinity QNM}).

\subsection{Asymptotic Iteration Method for the Dirac field}
\label{ss: aim Dirac}

To compute the QNFs of the Dirac field moving in the asymptotically adS 2D black hole (\ref{e: black hole two-dimensional}) we can take as a basis Eqs.\ (\ref{e: Dirac radial first}) and use the AIM for second order differential equations as explained in the previous section for the Klein-Gordon field. Nevertheless an alternative way can be used. As described in Ref.\ \cite{Cifti:2005 coupled} the AIM can be extended  to calculate the eigenvalues  for coupled systems of first order differential equations of the form
\begin{align} \label{e: aim coupled}
 \phi^\prime_1 = \Lambda_0 \phi_1 + S_0 \phi_2, \nonumber \\
 \phi^\prime_2 = \Omega_0 \phi_1 + P_0 \phi_2,
\end{align}
where $\Lambda_0$, $ S_0$, $\Omega_0 $, $P_0$ are functions of the independent variable $x$. 

In a similar way to the usual AIM previously described in Sect.\ \ref{ss: aim Klein-Gordon}, we notice that the first derivative of Eqs.\ (\ref{e: aim coupled}) simplifies to \cite{Cifti:2005 coupled}
\begin{align}
 \phi^{\prime \prime}_1 = \Lambda_1 \phi_1 + S_1 \phi_2, \nonumber \\
 \phi^{\prime \prime}_2 = \Omega_1 \phi_1 + P_1 \phi_2,
\end{align}
and we observe that the previous system of coupled differential equations is of the same form as the system (\ref{e: aim coupled}) with
\begin{align}
 \Lambda_1 &= \Lambda_0^\prime + \Lambda_0^2 + S_0 \Omega_0 ,  \nonumber \\
 S_1 &= S_0^\prime + \Lambda_0 S_0 + S_0 P_0 ,      \\
 \Omega_1  &= \Omega_0^\prime + \Lambda_0 \Omega_0 + P_0 \Omega_0 , \nonumber \\
 P_1 &= P_0^\prime + P_0^2 + S_0 \Omega_0. \nonumber 
\end{align}
As in the AIM for second order differential equations, we notice that the $(n+1)$-th derivative of the coupled system of first order differential equations (\ref{e: aim coupled}) takes the form \cite{Cifti:2005 coupled}
\begin{align}
 \phi^{(n+2)}_1 = \Lambda_{n+1} \phi_1 + S_{n+1} \phi_2, \nonumber \\
 \phi^{(n+2)}_2 = \Omega_{n+1} \phi_1 + P_{n+1} \phi_2,
\end{align}
where 
\begin{align} \label{e: recurrence relations coupled}
 \Lambda_{n+1} &= \Lambda_n^\prime + \Lambda_0 \Lambda_n + \Omega_0 S_n  ,\nonumber \\
 S_{n+1} &= S_n^\prime + P_0 S_n  + S_0  \Lambda_n  ,\nonumber \\
 \Omega_{n+1}  &= \Omega_n^\prime + \Lambda_0 \Omega_n + \Omega_0 P_n  ,\nonumber \\
 P_{n+1} &= P_n^\prime +  P_0 P_n  +  S_0 \Omega_n . 
\end{align}
In contrast to the AIM for second order differential equations, for which two recurrence relations are obtained (see the expressions (\ref{eq: recurrence general})), for a system of coupled first order differential equations we find four recurrence relations \cite{Cifti:2005 coupled}.

Furthermore, for the coupled system of first order differential equations, the asymptotic aspect of the AIM  demands that for sufficiently large $n$ \cite{Cifti:2005 coupled}
\begin{equation} \label{e: quantization coupled}
 \frac{S_{n+1}}{\Lambda_{n+1}}= \frac{S_{n}}{\Lambda_{n}} = \alpha,
\end{equation} 
which is similar to the expression (\ref{eq:condition 1}) of the AIM for second order differential equations, but notice that in both expressions the involved functions  satisfy different recurrence relations. From the previous formula we obtain that the QNFs are the stable numerical solutions of the equation \cite{Cifti:2005 coupled}
\begin{equation} \label{e: delta coupled}
 \Delta_n = \Lambda_{n} S_{n+1} - \Lambda_{n+1} S_{n} = 0.
\end{equation} 
For exactly solvable systems the previous quantization condition depends only of $\omega$, but in general depends on the independent variable and $\omega$ \cite{Cifti:2005 coupled}, hence to calculate the roots of  Eq.\ (\ref{e: delta coupled}), we evaluate it  in a convenient value of the independent variable.

In what follows we use this implementation of the AIM  for coupled systems of first order differential equations to calculate the QNFs of the Dirac field propagating in the 2D black hole (\ref{e: black hole two-dimensional}). As far as we know, this procedure is not used before to calculate the QNFs of the Dirac field. To this end we take as a basis the coupled system of equations (\ref{e: coupled radial odes}), but we notice that for the 2D black hole (\ref{e: black hole two-dimensional}) the functions $\mathcal{P}$ and $\mathcal{Q}$ defined in Appendix \ref{a: null dyad} fulfill $(\mathcal{P} \mathcal{Q} )^2 = 1$. Therefore, taking the radial functions $R_{l}$ in the form 
\begin{equation}
R_1  = \frac{(r-r_+)^{1/4}}{r^{3/4}} U_1, \qquad \qquad      R_2  = \frac{U_2}{(r-r_+)^{1/4} r^{1/4}} ,                                                                                                                                                                                                                                                                                                                                                                                                                                                                                                                                                                                                                                                                                                                                                                                                                                                                                                                                                                                                                                                                                                                                                                                      \end{equation} 
from Eqs.\ (\ref{e: coupled radial odes}) we obtain that the functions $U_1$, $U_2$ must be solutions of the coupled differential equations 
\begin{align} \label{e: coupled U}
 \frac{d U_1}{d r}  +\left( \frac{1/4}{r-r_+} - \frac{i \omega}{f} + \frac{1}{4 f}  \frac{d f}{dr} - \frac{3/4}{r}\right)U_1 &= - \frac{i m r^{1/2}}{\sqrt{f} (r-r_+)^{1/2}} U_2 , \\
 \frac{d U_2}{d r}  +\left( -\frac{1/4}{r-r_+} + \frac{i \omega}{f} + \frac{1}{4 f}  \frac{d f}{dr} - \frac{1/4}{r}\right)U_2 &=  \frac{i m (r-r_+)^{1/2} }{ \sqrt{f} r^{1/2} } U_1 . \nonumber
\end{align}

To impose  at the boundaries the required behavior  of the QNMs we propose that the functions $U_1$, $U_2$  are given by
\begin{equation}
 U_1 = h(r) V_1, \qquad \qquad  U_2 = h(r) V_2, 
\end{equation} 
where the function $h(r)$ is equal to
\begin{equation}
 h(r)= (r-r_+)^{-i \omega / 2 \kappa } r^{-m/2 + i \omega /( 2 \kappa) }.
\end{equation} 
Substituting into Eqs.\ (\ref{e: coupled U}) we find that the functions $V_1$, $V_2$ are solutions of 
\begin{align} \label{e: coupled V}
 \frac{d V_1}{d r} & +\left( \frac{1/4-i\omega/( 2 \kappa)}{r-r_+} - \frac{i \omega}{f} + \frac{1}{4 f}  \frac{d f}{dr} + \frac{i \omega / ( 2 \kappa)-m/2-3/4}{r}\right) V_1 \nonumber \\
 &= - \frac{i m r^{1/2}}{\sqrt{f} (r-r_+)^{1/2}} V_2 ,\nonumber \\
 \frac{d V_2}{d r} & +\left( -\frac{1/4 + i \omega / (2 \kappa )}{r-r_+} + \frac{i \omega}{f} + \frac{1}{4 f} \frac{d f}{dr} + \frac{i\omega/(2 \kappa)-m/2-1/4}{r}\right)V_2 \nonumber \\
 &=  \frac{i m (r-r_+)^{1/2} }{ \sqrt{f} r^{1/2} } V_1 . 
\end{align}

\begin{table}[ht]
\centering
\caption{First ten QNFs of the Klein-Gordon field produced by the HH method and by the AIM. We take $r_+=70$ and $m=1/10$.}
\begin{tabular}{|c|c|c|}
\hline
Mode number & HH method & AIM \\  \hline
0 & $0.078-0.136 i$ & $0.078-0.136 i$ \\ \hline
 1 & $0.095-0.168 i$ & $0.095-0.168 i$ \\ \hline
 2 & $0.112-0.200 i $& $0.112-0.200 i$ \\ \hline
 3 & $0.129-0.232 i$ & $0.129-0.232 i $\\ \hline
 4 & $0.147-0.265 i$ &$ 0.147-0.265 i$ \\  \hline
 5 & $0.165-0.297 i$ & $0.165-0.297 i$ \\ \hline
 6 & $0.182-0.329 i$ & $0.182-0.329 i$ \\ \hline
 7 & $0.200-0.362 i$ & $0.200-0.362 i $\\ \hline
 8 & $0.218-0.394 i$ & $0.218-0.394 i$ \\  \hline
 9 & $0.236-0.427 i$ &$ 0.236-0.427 i$ \\ \hline
\end{tabular}
\label{Tabla1}
\end{table}

\begin{figure}[ht]
\begin{center}
\includegraphics[scale=.9,clip=true]{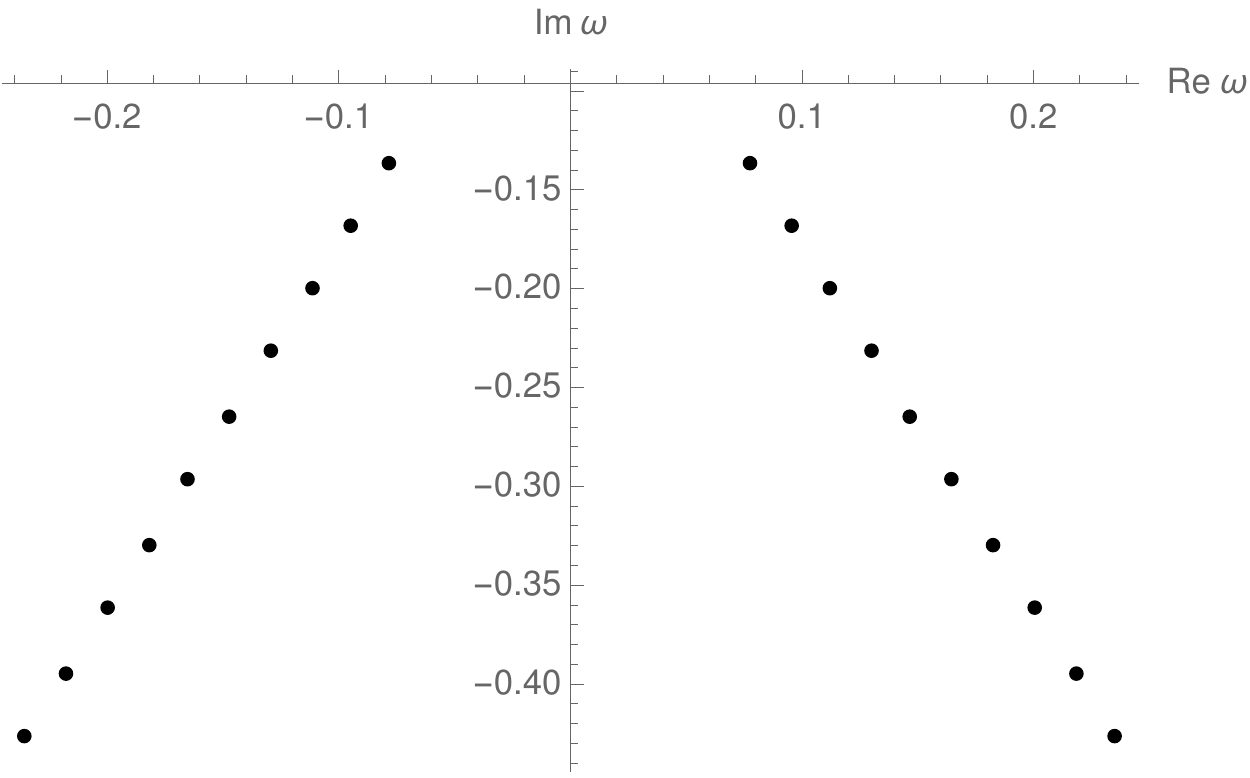}
\caption{First ten QNFs of the Klein-Gordon  field with mass $m=1/10$ propagating in the 2D black hole (\ref{e: black hole two-dimensional}) of radius $r_+=70$.  \label{f: kg r-70 m-10}} 
\end{center}
\end{figure}

\begin{figure}[ht]
\begin{center}
\includegraphics[scale=.9,clip=true]{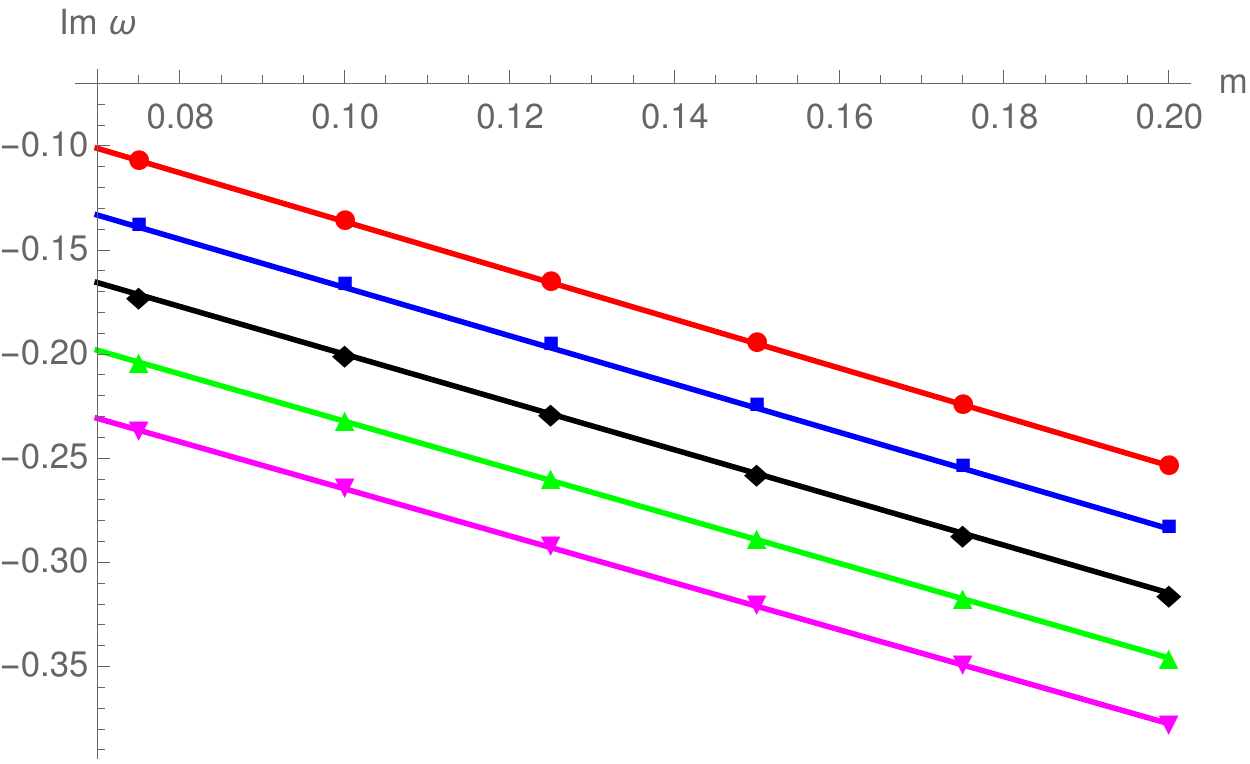}
\caption{For the first five QNFs of the Klein-Gordon field we show how the imaginary part depends on the mass of the field. We consider a 2D black hole  (\ref{e: black hole two-dimensional}) of radius $r_+=70$. \label{f: im vs m r-70 kg}}
\end{center}
\end{figure}

\begin{figure}[ht]
\begin{center}
\includegraphics[scale=.9,clip=true]{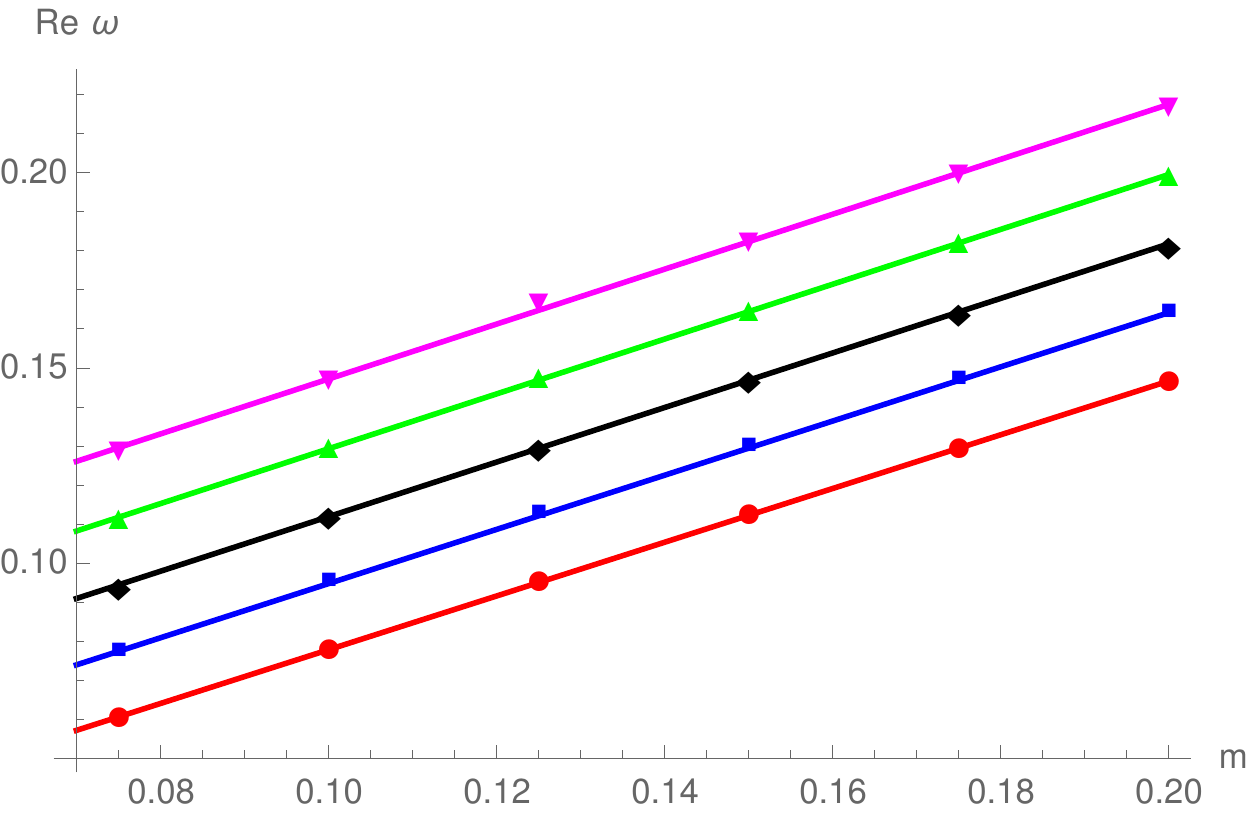}
\caption{For the first five QNFs of the Klein-Gordon field we show how the real part depends on the mass of the field. We consider a 2D black hole  (\ref{e: black hole two-dimensional}) of radius $r_+=70$. \label{f: re vs m r-70 kg}}
\end{center}
\end{figure}

\begin{table}[ht]
\centering
\caption{For the Klein-Gordon field we give the linear fits for the points shown in Figs.\ \ref{f: im vs m r-70 kg} and \ref{f: re vs m r-70 kg}. We take $r_+=70$.}
\begin{tabular}{|c|c|c|}
\hline
Mode number & Linear fit for $Im (\omega)$ vs $m$  &  Linear fit for $Re (\omega)$ vs $m$  \\ \hline
 0 & $-0.0192 - 1.1720 m$  & $0.0090 + 0.6880 m$  \\ \hline
 1 & $-0.0522 - 1.1586 m $ & $0.0254 + 0.6936 m$ \\ \hline
 2 & $-0.0855 - 1.1462 m $ & $0.0420 + 0.6985 m$ \\ \hline
 3 & $-0.1187 - 1.1363 m $ & $0.0591 + 0.7021 m$ \\ \hline
 4 & $-0.1520 - 1.1276 m $ & $0.0768 + 0.7030 m$  \\ \hline
\end{tabular} 
\label{Tabla-ajustes-kg}
\end{table}

\begin{figure}[ht]
\begin{center}
\includegraphics[scale=.9,clip=true]{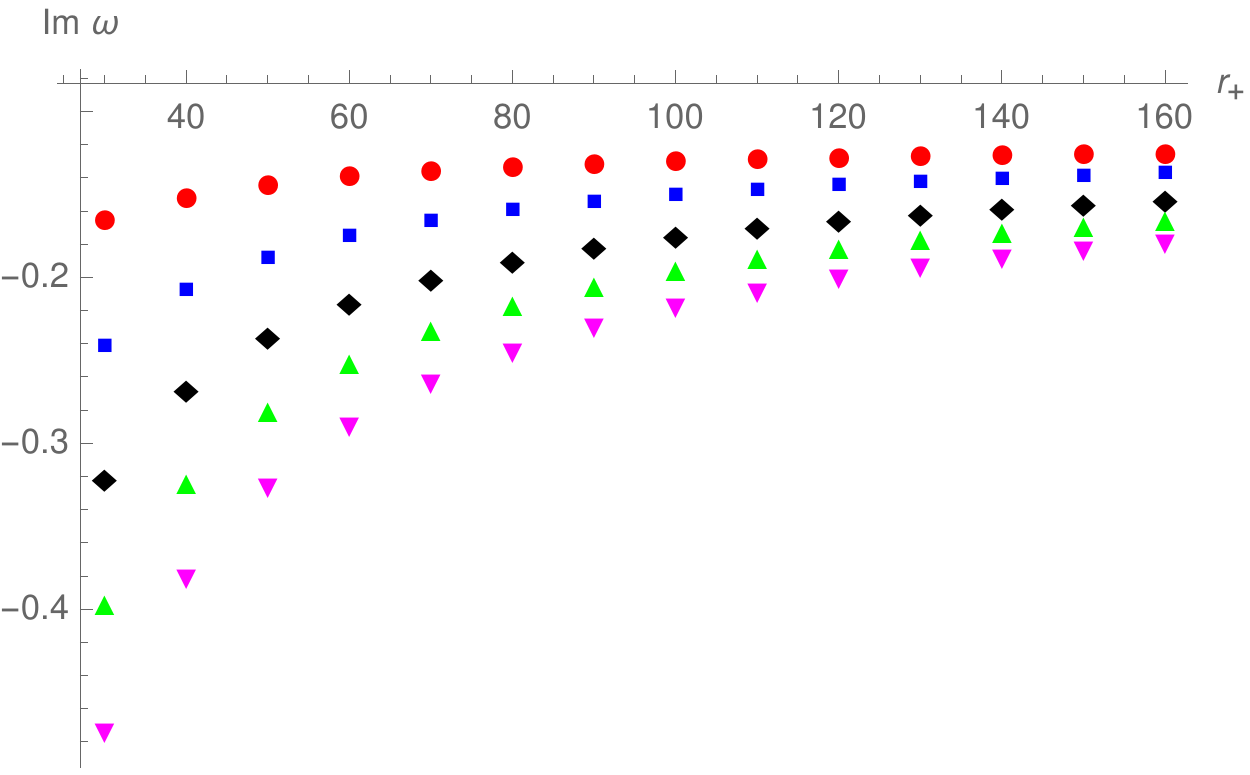}
\caption{For the first five QNFs of the Klein-Gordon field we show how the imaginary part depends on the radius of the 2D black hole (\ref{e: black hole two-dimensional}). We consider a Klein-Gordon field of mass $m=1/10$.  \label{f: im vs r m-10 kg}}
\end{center}
\end{figure}

\begin{figure}[ht]
\begin{center}
\includegraphics[scale=.9,clip=true]{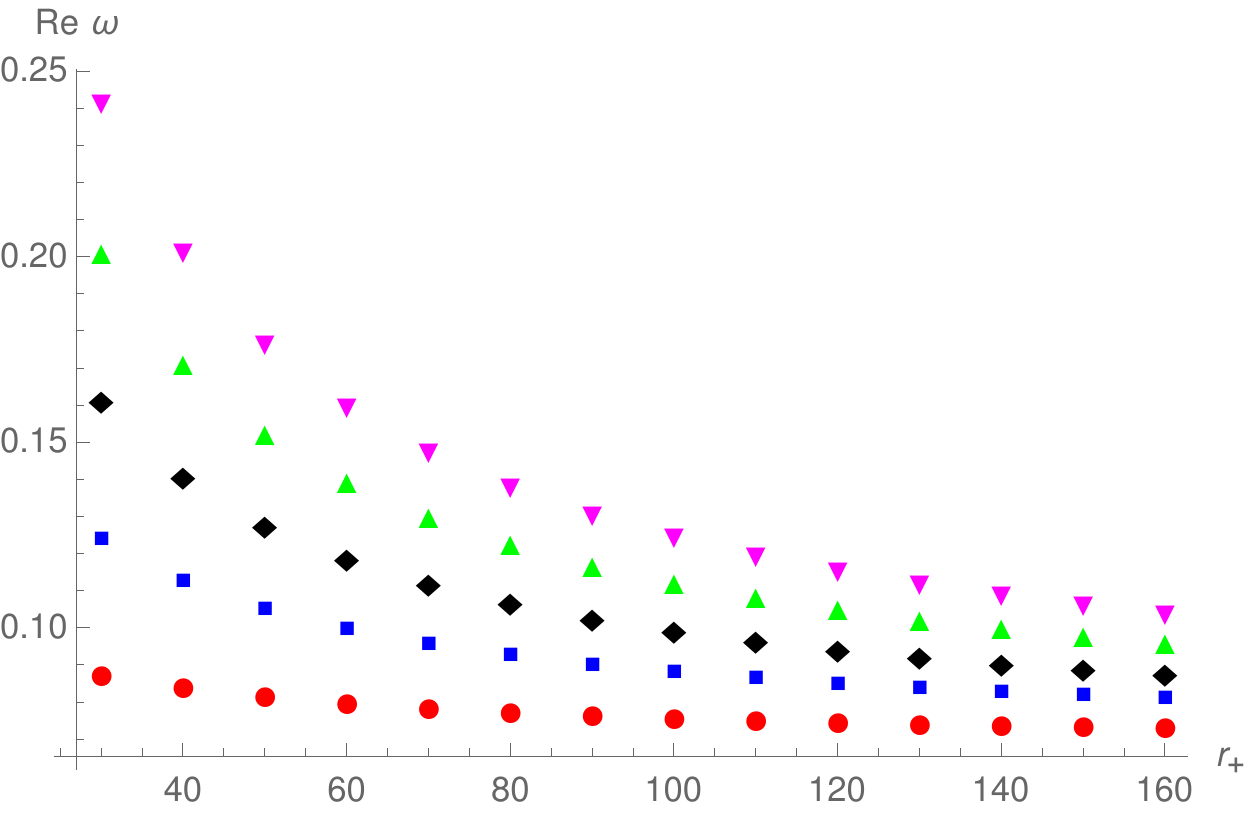}
\caption{For the first five QNFs of the Klein-Gordon field we show how the real part depends on the radius of the 2D black hole (\ref{e: black hole two-dimensional}). We consider a Klein-Gordon field of mass $m=1/10$. \label{f: re vs r m-10 kg}} 
\end{center}
\end{figure}

As we already  mentioned, the variable $r$ is defined in the semi-infinite interval $(r_+, \infty)$, but in the implementation of the AIM, it is convenient to have a finite range of the independent variable. Therefore we use the variable $z$ defined in the formula (\ref{e: z variable}) to write the previous coupled differential equations in the form 
\begin{align} \label{e: coupled V z variable}
 \frac{d V_1}{d z} & = \left( - \frac{i \omega r_+}{1-z^3} + \frac{2+z^3}{4(1-z^3)z}+ \frac{1/4-i\omega/( 2 \kappa)}{z(1-z)}   + \frac{i \omega / ( 2 \kappa)-m/2-3/4}{z}\right) V_1 \nonumber \\
 &+  \frac{i m r_+}{(1-z) (1+z+z^2)^{1/2} z} V_2 ,\nonumber \\
 \frac{d V_2}{d z} & = \left(  \frac{i \omega r_+}{1-z^3} + \frac{2+z^3}{4(1-z^3)z}- \frac{1/4+i\omega/( 2 \kappa)}{z(1-z)}   + \frac{i \omega / ( 2 \kappa)-m/2-1/4}{z}\right) V_2 \nonumber \\
 &-  \frac{i m r_+}{ (1+z+z^2)^{1/2} z} V_1 .
\end{align}
This coupled system is of the form (\ref{e: aim coupled}) and we can identify the functions $\Lambda_0$, $S_0$, $\Omega_0$, and $P_0$ as follows
\begin{align} \label{e: lambda s omega p coupled}
 \Lambda_0 (z) &= - \frac{i \omega r_+}{1-z^3} + \frac{2+z^3}{4(1-z^3)z}+ \frac{1/4-i\omega/( 2 \kappa)}{z(1-z)}   + \frac{i \omega / ( 2 \kappa)-m/2-3/4}{z} ,\nonumber \\
 S_0 (z) &=  \frac{i m r_+}{(1-z) (1+z+z^2)^{1/2} z}, \nonumber \\
 \Omega_0 (z) &= -  \frac{i m r_+}{ (1+z+z^2)^{1/2} z} ,\\
 P_0 (z) &=  \frac{i \omega r_+}{1-z^3} + \frac{2+z^3}{4(1-z^3)z}- \frac{1/4+i\omega/( 2 \kappa)}{z(1-z)}   + \frac{i \omega / ( 2 \kappa)-m/2-1/4}{z}. \nonumber
\end{align}
From these quantities and taking into account the recurrence relations (\ref{e: recurrence relations coupled}) we calculate the quantities $\Delta_n$ of Eq.\ (\ref{e: delta coupled}), whose stable roots are the QNFs of the Dirac field in the 2D black hole (\ref{e: black hole two-dimensional}). Also we notice that in the functions $S_0$, $\Omega_0$ appear a factor that includes a square root, but this factor is well defined at the boundaries $z=0$ and $z=1$.

\section{Results}

In what follows we describe our numerical results for the QNFs of the Klein--Gordon field and of the Dirac field.

\subsection{Numerical results for the Klein-Gordon field}
\label{ss: numerical Klein-Gordon}

For a specific configuration, in Table \ref{Tabla1} we show the values of the QNFs for the Klein-Gordon field that produce the HH method and the AIM. For the first ten QNMs, we observe that the two numerical methods yield QNFs which agree to three decimal places. We point out that  for all the examples studied in the present work the two methods produce values of the QNFs which agree to three decimal places (see Appendix \ref{a: additional tables} where we present some numerical results with more decimal places). We also notice that the two numerical methods yield stable QNFs, that is, frequencies with negative imaginary part.\footnote{In this work our convention for the QNFs is $\omega= \textrm{Re}(\omega) + i \textrm{Im}(\omega)$. Therefore a complex frequency with negative imaginary part is related with a wave that decays in time (see the time dependence in the formula (\ref{e: ansatz Klein-Gordon})).}

In Fig.\ \ref{f: kg r-70 m-10} we plot the real part and the imaginary part of the first ten QNFs for the Klein-Gordon field of mass $m=1/10$ propagating in the 2D black hole (\ref{e: black hole two-dimensional}) with radius of the horizon equal to $r_+ = 70$. We notice that the graph $Im(\omega)\,\,\, vs \,\,\, Re(\omega)$ shows a linear relation between the imaginary part and the real part of the QNFs for the Klein-Gordon field as we change the mode number.   We notice that in  Fig.\ \ref{f: kg r-70 m-10} appear twenty points, but we consider as equivalent the QNFs  with the same imaginary part, that is, we find two QNFs with the same imaginary part, but the real part of the first is the negative of the real part of the second.

We notice that in the Figures of this paper (except for Figs.\ \ref{f: kg r-70 m-10}, \ref{f: dirac r-70 m-10}) we label the modes of the field as follows: $n=0$, red circle; $n=1$, blue square; $n=2$, black diamond; $n=3$, green triangle; $n=4$, inverted magenta triangle.

For $r_+ = 70$, in Figs.\ \ref{f: im vs m r-70 kg} and \ref{f: re vs m r-70 kg} we present the variation of the imaginary part and the real part of the first five QNFs when we modify the mass of the Klein-Gordon field.  In these figures we observe that the imaginary part and the real part of the QNFs change in a linear way as the mass varies. In Figs.\ \ref{f: im vs m r-70 kg} and \ref{f: re vs m r-70 kg} we observe that the imaginary part decreases, whereas the real part increases as the mass of the field increases, thus the QNMs of the Klein-Gordon decay faster and make more oscillations per time unit as the mass of the Klein-Gordon field increases. 

 Furthermore, in Figs.\ \ref{f: im vs m r-70 kg} and \ref{f: re vs m r-70 kg}  we show the plots of the linear fits given in Table \ref{Tabla-ajustes-kg}. From the linear fits of Table \ref{Tabla-ajustes-kg} we see that for the plots $Im(\omega)$ vs $m$ the absolute value of the slope decreases as the mode number increases. For the plots  $Re(\omega)$ vs $m$ we notice that the slope increases as the mode number increases. In both cases the change of slope with the mode number is small.

In Figs.\ \ref{f: im vs r m-10 kg} and \ref{f: re vs r m-10 kg} we draw the behavior of the imaginary part and of the real part for the first five QNFs when the radius of the horizon varies and the mass of the Klein-Gordon field is $m=1/10$. In these figures we see that the imaginary part increases, whereas the real part decreases as the radius of the black hole increases, thus, for a given mode the decay time increases and the oscillation frequency  decreases as the horizon radius increases. In both figures we observe that the imaginary part and the real part of the fifth QNFs have the larger variations, whereas the imaginary part and the real part of the fundamental frequency show the smaller changes as the horizon radius increases. 

In Figs.\ \ref{f: im vs r m-10 kg} and \ref{f: re vs r m-10 kg} we also notice  that the QNFs of the Klein-Gordon field in the 2D black hole (\ref{e: black hole two-dimensional}) are more compactly distributed in the $\omega$ plane for bigger values of the horizon radius.

\subsection{Numerical results for the Dirac field}
\label{ss: numerical Dirac}

\begin{table}[ht]
\centering
\caption{First ten QNFs of the Dirac field produced by the HH method and by the AIM for coupled systems of first order differential equations. We take $r_+=70$ and $m=1/10$.}
\begin{tabular}{|c|c|c|}
\hline
Mode number  & HH method & AIM coupled \\  \hline
0 & $0.078-0.135 i$ & $0.078-0.135 i$ \\ \hline
 1 & $0.095-0.166 i$ & $0.095-0.166 i$ \\ \hline
 2 & $0.112-0.198 i$ & $0.112-0.198 i$ \\ \hline
 3 & $0.130-0.230 i$ & $0.130-0.230 i$ \\ \hline
 4 & $0.148-0.262 i$ & $0.148-0.262 i$ \\ \hline
 5 & $0.166-0.294 i$ & $0.166-0.294 i$ \\ \hline
 6 &$ 0.184-0.326 i$ & $0.184-0.326 i$ \\ \hline
 7 &$ 0.202-0.358 i$ & $0.202-0.358 i$ \\ \hline
 8 & $0.220-0.390 i$ & $0.220-0.390 i$ \\ \hline
 9 & $0.238-0.423 i$ & $0.238-0.423 i$ \\ \hline
\end{tabular}
\label{Tabla2}
\end{table}

\begin{table}[ht]
\centering
\caption{ We show the QNFs of the Dirac field produced by the effective potentials $V_1$ and $V_2$. We get these values for the QNFs using the HH method.  We take  $r_+ = 70$ and $m = 1/10$.}
\begin{tabular}{|c|c|c|}
\hline
Mode number & $V_2$ & $V_1$ \\  \hline
0 & $0.078-0.135 i$ & $0.078-0.135 i$ \\ \hline
1 & $0.095-0.166 i$ & $0.095-0.166 i$ \\ \hline
2 & $0.112-0.198 i$ & $0.112-0.198 i$ \\ \hline
3 & $0.130-0.230 i$ & $0.130-0.230 i$ \\ \hline
4 & $0.148-0.262 i$ & $0.148-0.262 i$ \\ \hline
5 & $0.166-0.294 i$ & $0.166-0.294 i$ \\ \hline
6 & $0.184-0.326 i$ & $0.184-0.326 i$ \\ \hline
7 & $0.202-0.358 i$ & $0.202-0.358 i$ \\ \hline
8 & $0.220-0.390 i$ & $0.220-0.390 i$ \\ \hline
9 & $0.238-0.423 i$ & $0.238-0.423 i$ \\ \hline
\end{tabular} 
\label{Tabla-V1+V2}
\end{table}

\begin{figure}[ht]
\begin{center}
\includegraphics[scale=.9,clip=true]{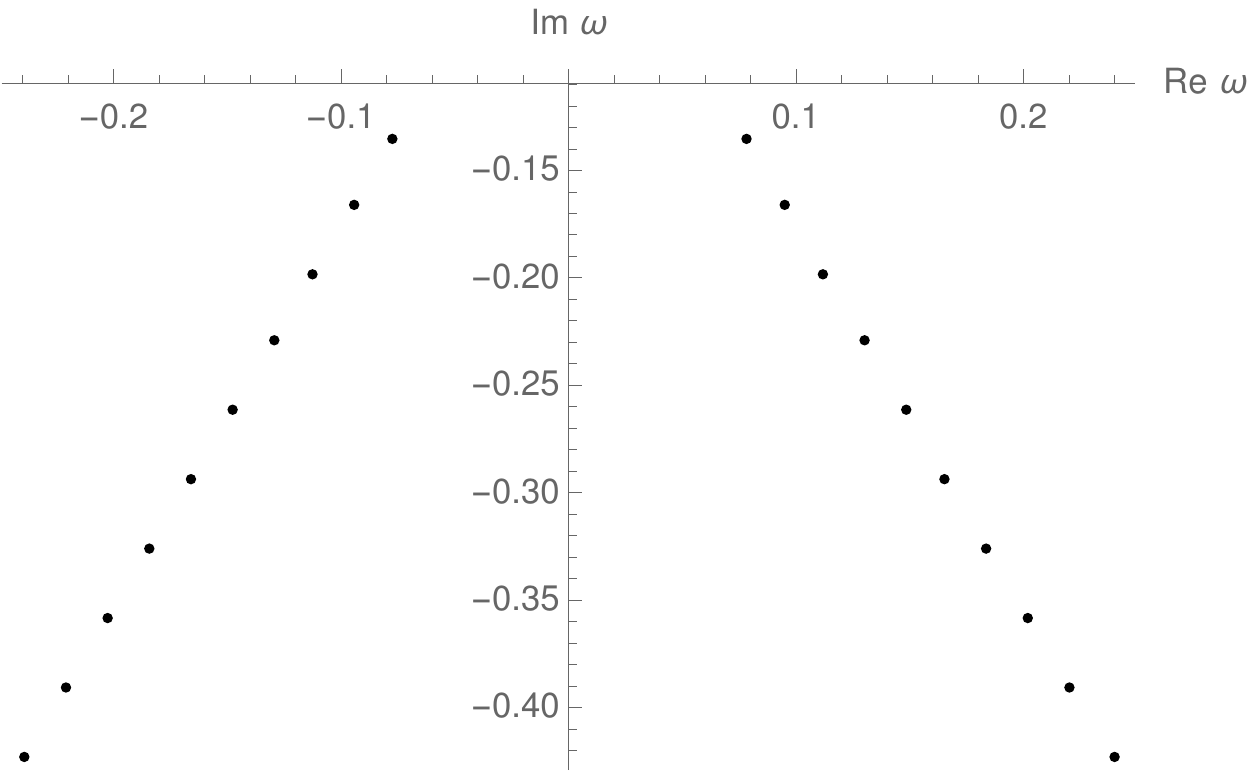}
\caption{First ten QNFs of the Dirac  field with mass $m=1/10$ propagating in the 2D black hole (\ref{e: black hole two-dimensional}) of radius $r_+=70$.  \label{f: dirac r-70 m-10}} 
\end{center}
\end{figure}

\begin{figure}[ht]
\begin{center}
\includegraphics[scale=.9,clip=true]{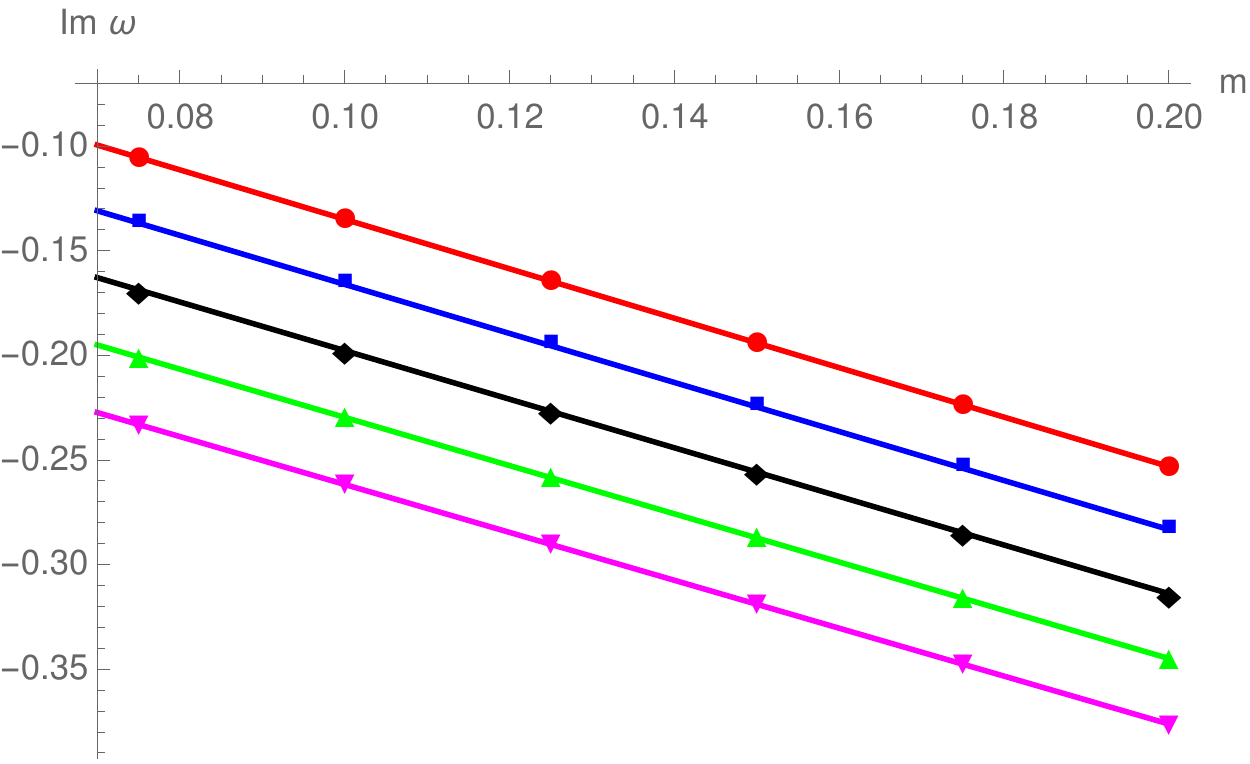}
\caption{For the first five QNFs of the Dirac field we show how the imaginary part depends on the mass of the field. We consider a 2D black hole  (\ref{e: black hole two-dimensional}) of radius $r_+=70$. \label{f: im vs m r-70 dirac}} 
\end{center}
\end{figure}

\begin{figure}[ht]
\begin{center}
\includegraphics[scale=.9,clip=true]{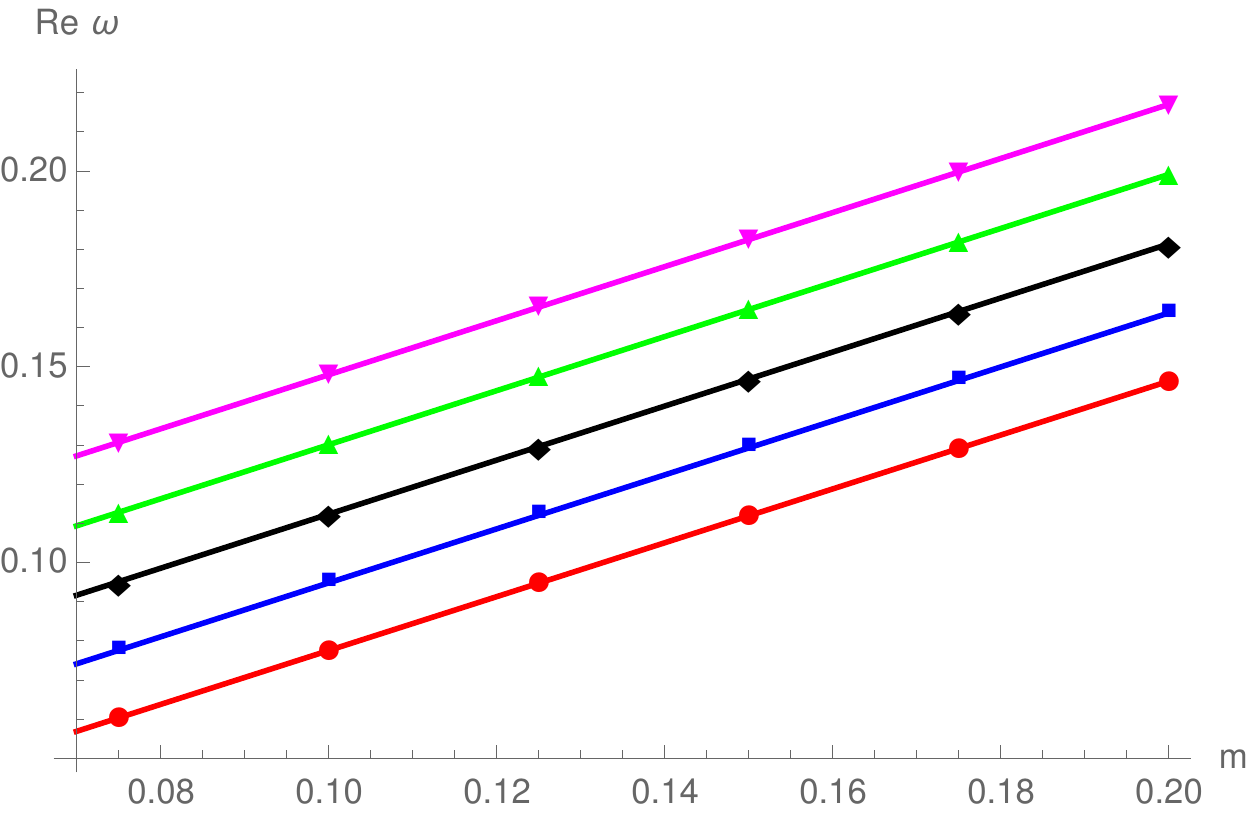}
\caption{For the first five QNFs of the Dirac field we show how the real part depends on the mass of the field. We consider a 2D black hole  (\ref{e: black hole two-dimensional}) of radius $r_+=70$. \label{f: re vs m r-70 dirac}} 
\end{center}
\end{figure}

\begin{table}[ht]
\centering
\caption{For the Dirac field we give the linear fits for the points shown in Figs.\ \ref{f: im vs m r-70 dirac} and \ref{f: re vs m r-70 dirac}. We take $r_+=70$.}
\begin{tabular}{|c|c|c|}
\hline
Mode number & Linear fit for $Im (\omega)$ vs $m$  &  Linear fit for $Re (\omega)$ vs $m$  \\ \hline
 0 & $-0.0168 - 1.1818 m$  & $ 0.0087 + 0.6876 m$  \\ \hline
 1 & $-0.0489 - 1.1717m $ & $ 0.0259 + 0.6890 m$ \\ \hline
 2 & $-0.0816 - 1.1614 m $ & $ 0.0433 + 0.6898 m$ \\ \hline
 3 & $-0.1144 - 1.1524 m $ & $ 0.0610 + 0.6901 m$ \\ \hline
 4 & $-0.1472 - 1.1450 m $ & $ 0.0789 + 0.6902 m$  \\ \hline
\end{tabular} 
\label{Tabla-ajustes-dirac}
\end{table}

\begin{figure}[ht]
\begin{center}
\includegraphics[scale=.9,clip=true]{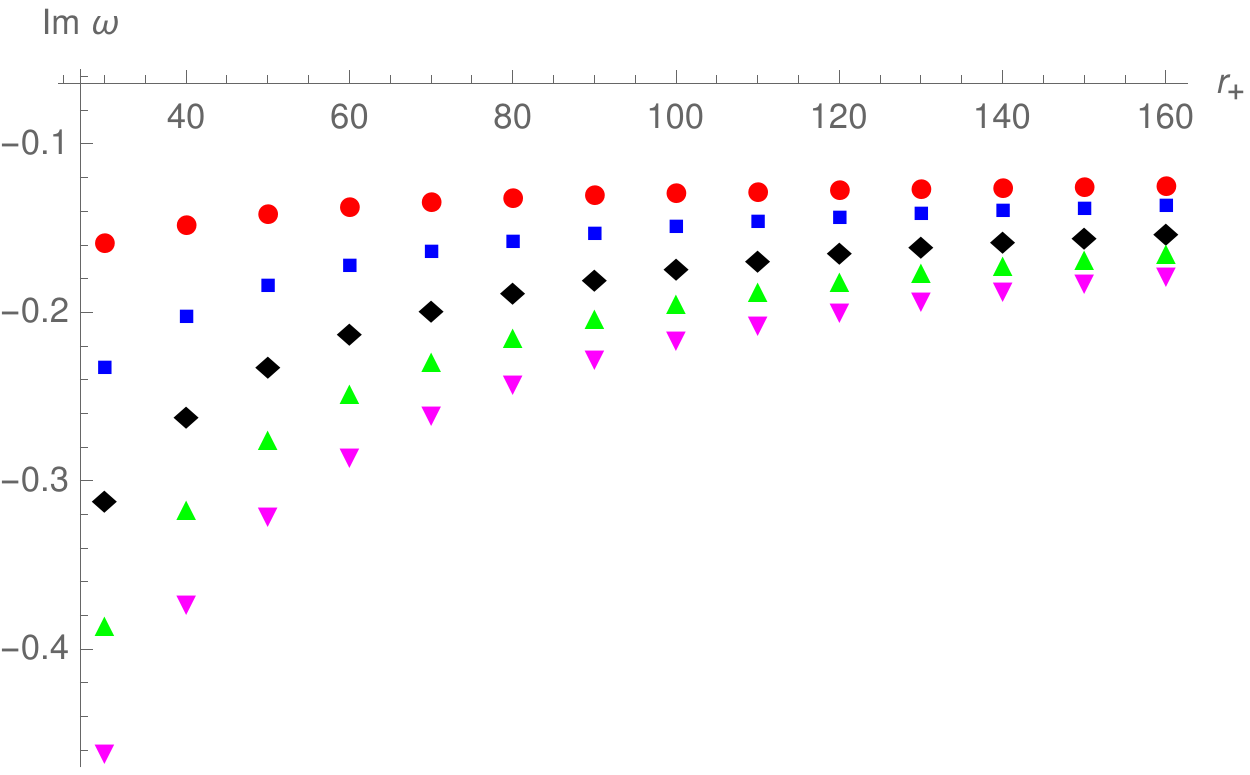}
\caption{For the first five QNFs of the Dirac field we show how the imaginary part depends on the radius of the 2D black hole (\ref{e: black hole two-dimensional}). We consider a Dirac field of mass $m=1/10$. \label{f: im vs r m-10 dirac}} 
\end{center}
\end{figure}

\begin{figure}[ht]
\begin{center}
\includegraphics[scale=.9,clip=true]{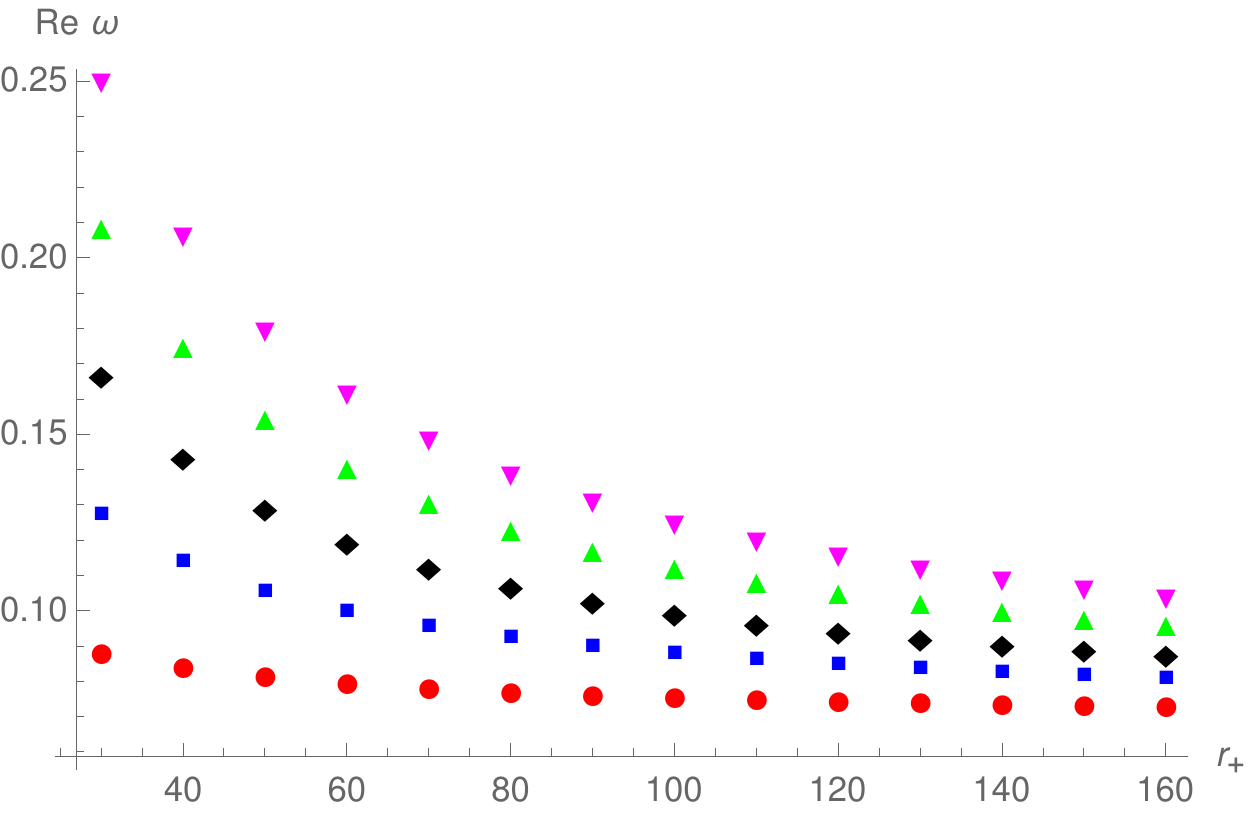}
\caption{For the first five QNFs of the Dirac field we show how the real part depends on the radius of the 2D black hole (\ref{e: black hole two-dimensional}). We consider a Dirac field of mass $m=1/10$. \label{f: re vs r m-10 dirac}} 
\end{center}
\end{figure}

For the Dirac field propagating in the 2D black hole (\ref{e: black hole two-dimensional}) we find that,  for the first ten modes,  the two methods produce  QNFs that   agree to three decimal places. This fact is observed in Table \ref{Tabla2} for a specific configuration (see  Appendix \ref{a: additional tables} for some numerical results with more decimal places). Furthermore, both methods yield QNFs with negative imaginary part, hence the QNMs of the Dirac field are stable in the 2D black hole (\ref{e: black hole two-dimensional}).

Taking as a basis the HH method, for $r_+ =70 $ and $m= 1/10 $, in Table \ref{Tabla-V1+V2} we observe that the QNFs produced by the effective potentials $V_1$ and $V_2$ of Eq.\ (\ref{e: Potentials Dirac 1/2})  are equal. Also, see the end of Appendix \ref{a: null dyad} for another argument pointing out that these effective potentials generate the same QNFs. At present time we  only have evidence that the effective potentials of Eq.\ (\ref{e: Potentials Dirac 1/2}) produce the same spectrum of QNF in the example studied in this work, but we think it would be convenient to show a more general result or find a case where they produce a different spectrum of QNF.

For the Dirac field, in Figs.\ \ref{f: dirac r-70 m-10}--\ref{f: re vs r m-10 dirac} we observe that the behavior of its QNFs is similar to that previously obtained  for the QNFs of the Klein-Gordon field, hence we describe briefly the results for the Dirac field.

\begin{itemize}

\item For the QNFs of the Dirac field, in Fig.\ \ref{f: dirac r-70 m-10} we see that the relation between the imaginary part and the real part of  the QNFs is linear as we change the mode number, in a similar way to the QNFs of the Klein-Gordon field.   
 
\item In Figs.\ \ref{f: im vs m r-70 dirac} and \ref{f: re vs m r-70 dirac} we observe that the imaginary part and the real part of the QNFs for the Dirac field change in a linear way when we modify the value of the field's mass. The linear fits of the data given in Figs.\ \ref{f: im vs m r-70 dirac} and \ref{f: re vs m r-70 dirac} are written in Table \ref{Tabla-ajustes-dirac}. In a similar way to the QNFs of the Klein-Gordon field we see that the imaginary part of the QNFs decreases as the field's mass increases, whereas the real part of the QNFs increases as the field's mass increases. As for the Klein-Gordon field, for the plots $Im(\omega)$ vs $m$ the absolute value of the slope decreases as the mode number increases. For the plots $Re(\omega)$ vs $m$ we find that the slope increases as the mode number increases. We also see that in both cases,	 the changes of the slope with the mode number are small.
 
\item When we change the radius of the horizon, in Figs.\ \ref{f: im vs r m-10 dirac} and \ref{f: re vs r m-10 dirac} we notice that for the QNFs of the Dirac field the imaginary part increases as the horizon radius increases, while the real part decreases as the horizon radius increases. These behaviors are similar to those for the QNFs of the Klein-Gordon field.
 
\end{itemize}

As for the Klein-Gordon field, we also note that the QNFs of the Dirac field are more compactly distributed in the complex $\omega$ plane when the horizon radius is bigger.

\section{Discussion}
\label{s: Discussion}

As far as we know, the AIM for coupled systems of first order differential equations has not previously used to calculate the QNFs of black holes. We find that the AIM for coupled systems of first order equations yields QNFs of the Dirac field in agreement with those that produces the HH method (see also Appendices \ref{a: KG coupled AIM} and \ref{a: iaim coupled}). We believe that this method is appropriate for computing the QNFs of the Dirac field, since in many spacetimes its equation of motion simplifies to a coupled system of first order differential equations \cite{LopezOrtega:2009qc}, \cite{Gibbons:1993hg}, \cite{Cotaescu:1998ay}, and in general the proposed method may be helpful to compute  the QNFs of classical fields whose equations of motion simplify to a pair of coupled first order differential equations. Also this method for calculating the QNFs of the Dirac field may be useful to verify the results that are obtained with other numerical procedures. We notice that in Appendix \ref{a: iaim coupled}, taking as a basis the improved formulation of the AIM for second order differential equations \cite{Cho:2009cj}, \cite{Cho:2011sf}, we present an improved formulation of the AIM for coupled systems of first order differential equations. 

For the two fields propagating in the  asymptotically adS 2D black hole (\ref{e: black hole two-dimensional}), we find that the HH method and the AIM produce the same values for their first ten QNFs (see also Appendices \ref{a: KG coupled AIM}, \ref{a: iaim coupled}). Furthermore, in the numerical results we obtain QNFs with negative imaginary part, therefore the asymptotically adS 2D black hole (\ref{e: black hole two-dimensional}) is stable under the Klein-Gordon and Dirac perturbations. A relevant fact is that its QNFs are complex, in contrast to other asymptotically adS 2D black hole in which purely imaginary QNFs are found \cite{Cordero:2012je}. As discussed in Sects.\ \ref{ss: numerical Klein-Gordon} and \ref{ss: numerical Dirac}, in the numerical results we observe that  the QNFs of the Klein-Gordon and Dirac fields behave in a similar way as we change the field's mass or the radius of the horizon. As the mass increases, both fields decay faster and the oscillation frequencies increase. Furthermore, as the radius of the horizon increases the decay time increases, whereas the oscillation frequency decreases.  

From the values presented in Tables \ref{Tabla1} and \ref{Tabla2} we see that the QNFs of the Klein-Gordon and  Dirac fields are similar,  even though one field is bosonic and the other field is fermionic. Furthermore, in Table \ref{Tabla3} we display the numerical values of the QNFs for the Klein-Gordon and Dirac fields propagating in the 2D black holes (\ref{e: black hole two-dimensional}) with radii $r_+=30$ and $r_+=150$. In this table we observe that for $r_+=150$ the QNFs of the two fields are essentially identical, also we note that for $r_+=30$ their QNFs are similar but they show more differences than the frequencies of the 2D black holes with radii $r_+=70$ and $r_+=150$. We also notice that for $r_+=30$ the fundamental QNFs are closer in value than the tenth QNFs.  Thus the numerical results show that when the horizon radius increases the QNFs of the two  fields are more similar. From the viewpoint of the Schr\"odinger type equations, these results are intriguing,  because the analysis of the problem shows that for the two fields their effective potentials are very different. We also note that the effective potentials for the Dirac field fulfill
\begin{equation} \label{e: potentials relation}
 V_{l} = V_{KG} \mp \frac{i \omega}{2}\frac{d f}{d r} - \frac{f}{4}\frac{d ^{2} f}{d r^{2}} + \frac{1}{16}\left(\frac{d f}{d r}\right)^{2} .
\end{equation} 
Although these differences, for some values of the physical parameters, the effective potentials $V_{l} $, $V_{KG}$ produce almost isospectral QNFs in the asymptotically adS 2D black hole (\ref{e: black hole two-dimensional}). Furthermore, as we previously mentioned, the QNFs of the effective potentials $V_{l} $, $V_{KG}$ behave in a similar way when we change the physical parameters. Thus our numerical results indicate that for the 2D black hole (\ref{e: black hole two-dimensional}) the last terms in the right hand side of the formula (\ref{e: potentials relation}) have a small contribution to the value of the QNFs for the Dirac field. To start understanding this fact we notice the following facts.  Near the horizon the last three terms in the formula (\ref{e: potentials relation}) only produce a shift in the effective frequency that appears in the Schr\"odinger type equations, from $\omega$ for the Klein-Gordon field to $(\omega \pm i \kappa/2)$ for the Dirac field. Therefore, near the horizon, for the Klein-Gordon field we get  the radial solutions (\ref{e: near horizon Klein})  whereas for the Dirac field we obtain the solutions (\ref{e: near horizon Dirac}). This fact does not prevent us from choosing the relevant solution for the QNMs. As $r \to \infty$, the last three terms in Eq.\ (\ref{e: potentials relation}) produce that the dominant behavior of the effective potentials change from $V_{KG} \sim  a^2 r^2 m^2$  for the Klein-Gordon field to $V_l \sim a^2 r^2 (m^2 -a^2/4)$ for the Dirac field, that is, only produce a shift in the value of the mass. In our case this change does not prevent us from choosing the appropriate solution that satisfies the boundary condition of the QNMs at the asymptotic region.

\begin{table}[ht]
\centering
\caption{For the 2D black holes (\ref{e: black hole two-dimensional}) with radii $r_+=30$ and $r_+=150$ we display the QNFs of the Klein-Gordon and Dirac fields of mass $m=1/10$. }
\begin{tabular}{|p{2.1cm} |p{2.1cm}|p{2cm}|p{2.1cm}|p{2cm}|  }
 \hline
 \multicolumn{1}{|c|}{} & \multicolumn{2}{|c|}{$r_+=30$} & \multicolumn{2}{|c|}{$r_+=150$} \\ \hline
Mode number & Klein-Gordon & Dirac & Klein-Gordon & Dirac  \\\hline
 0 & $0.087-0.166 i$ & $0.087-0.160 i$ & $0.073-0.126 i $ & $0.073-0.126 i$ \\\hline
 1 & $0.123-0.243 i$ & $0.127-0.235 i$ & $0.081-0.141 i$ & $0.081-0.140 i$ \\\hline
 2 & $0.161-0.321 i$ & $0.167-0.311 i$ & $0.089-0.155 i$ & $0.089-0.154 i$ \\\hline
 3 & $0.201-0.398 i$ & $0.208-0.387 i$ & $0.097-0.169 i$ & $0.097-0.169 i$ \\\hline
 4 & $0.241-0.474 i$ & $0.250-0.462 i$ & $0.106-0.184 i$ & $0.106-0.183 i$ \\\hline
 5 & $0.281-0.551 i$ & $0.291-0.538 i$ & $0.114-0.199 i$ & $0.114-0.198 i$ \\\hline
 6 & $0.322-0.627 i$ & $0.333-0.614 i$ & $0.122-0.214 i$ & $0.122-0.213 i$ \\\hline
 7 & $0.364-0.704 i$ & $0.376-0.689 i$ & $0.131-0.229 i$ & $0.131-0.228 i$ \\\hline
 8 & $0.405-0.780 i$ & $0.418-0.765 i$ & $0.139-0.244 i$ & $0.139-0.243 i$ \\\hline
 9 & $0.447-0.856 i$ & $0.460-0.841 i$ & $0.148-0.258 i$ & $0.148-0.257 i $\\\hline
\end{tabular}
\label{Tabla3}
\end{table}

\section{Acknowledgments}

This work was supported by CONACYT M\'exico, SNI M\'exico, EDI-IPN, COFAA-IPN, and Research Project IPN SIP 20210485.

\appendix

\section{Additional tables\footnote{We thank one of the reviewers for suggesting us to make the comparison shown in Tables \ref{Tabla-KG-comp} and \ref{Tabla-Dirac-comp}.}}
\label{a: additional tables}

For the QNFs of the  Klein-Gordon field, in Table \ref{Tabla-KG-comp}  we present the values that produce the HH method and the AIM to higher digits (compare with Table \ref{Tabla1}). In Table \ref{Tabla-KG-comp} we observe that for the first seven QNMs both numerical methods produce the same QNFs to seven decimal places, but for the last three QNFs we notice some differences in the last digits.

\begin{table}[ht]
\centering
\caption{For the Klein-Gordon field we show its QNFs to higher digits. The QNFs are produced by the HH method and the AIM. We take  $r_+ = 70$ and $m = 1/10$.}
\begin{tabular}{|c|c|c|}
\hline
Mode number & HH method & AIM \\  \hline
 0 & $0.0778658-0.1362857 i$ & $0.0778658-0.1362857 i$ \\ \hline
 1 & $0.0948790-0.1678734 i$ & $0.0948790-0.1678734 i$ \\ \hline
 2 & $0.1120717-0.1999259 i$ & $0.1120717-0.1999259 i$ \\ \hline
 3 & $0.1294365-0.2322016 i$ & $0.1294365-0.2322016 i$ \\ \hline
 4 & $0.1469485-0.2645830 i$ & $0.1469485-0.2645830 i$ \\ \hline
 5 & $0.1645814-0.2970118 i$ & $0.1645814-0.2970118 i$ \\ \hline
 6 & $0.1823128-0.3294582 i$ & $0.1823128-0.3294582 i$ \\ \hline
 7 & $0.2001263-0.3619064 i$ & $0.2001251-0.3619070 i$ \\ \hline
 8 & $0.2180128-0.3943701 i$ & $0.2180044-0.3943504 i$ \\ \hline
 9 & $0.2357184-0.4268551 i$ & $0.2359398-0.4267845 i$ \\ \hline
\end{tabular} 
\label{Tabla-KG-comp}
\end{table}

In a similar way, in Table \ref{Tabla-Dirac-comp} we give the QNFs of the Dirac field produced by the HH method and the AIM to higher digits (compare with Table \ref{Tabla2}). As for the Klein-Gordon field, the first seven QNFs of the Dirac field coincide to seven decimal places and the last three QNFs show differences in the last digits.

\begin{table}[ht]
\centering
\caption{For the Dirac field we show its QNFs to higher digits. The QNFs are produced by the HH method and the AIM. We take  $r_+ = 70$ and $m = 1/10$.}
\begin{tabular}{|c|c|c|}
\hline
Mode number & HH method & AIM \\  \hline
 0 & $0.0775319-0.1349311 i$ & $0.0775319-0.1349311 i$ \\ \hline 
 1 & $0.0948471-0.1659988 i$ & $0.0948471-0.1659988 i$ \\ \hline
 2 & $0.1124024-0.1976164 i$ & $0.1124024-0.1976164 i$ \\ \hline 
 3 & $0.1301370-0.2295252 i$ & $0.1301370-0.2295252 i$ \\ \hline
 4 & $0.1480059-0.2615908 i$ & $0.1480059-0.2615908 i$ \\ \hline
 5 & $0.1659765-0.2937427 i$ & $0.1659765-0.2937427 i$ \\ \hline
 6 & $0.1840250-0.3259428 i$ & $0.1840250-0.3259428 i$ \\ \hline
 7 & $0.2021342-0.3581700 i$ & $0.2021343-0.3581696 i$ \\ \hline
 8 & $0.2202862-0.3904068 i$ & $0.2202922-0.3904111 i$ \\ \hline
 9 & $0.2385630-0.4225911 i$ & $0.2384893-0.4226599 i$ \\ \hline
\end{tabular} 
\label{Tabla-Dirac-comp}
\end{table}

\section{The Dirac equation in a null dyad}
\label{a: null dyad}

Reference \cite{LopezOrtega:2011sc} shows that in a static 2D spacetime with diagonal metric and in the chiral representation for the gamma matrices,  the Dirac equation (\ref{e: Dirac equation})  simplifies to a pair of Schr\"odinger type equations with effective potentials involving square roots of the metric functions (see the expressions (\ref{e: potentials Dirac square root})). Motivated by the Newman-Penrose formalism of four-dimensional spacetimes \cite{NP-article-JMP}, for a static and diagonal  2D spacetime in what follows we use a null dyad to simplify the Dirac equation   to a different pair of Schr\"odinger type equations with potentials not including square roots. 

It is convenient to write the metric of the 2D spacetime under study as
\begin{equation} \label{e: metric bidi PQ}
d s^{2} = \mathcal{P}^{2}d t^{2} - \mathcal{Q}^{2}d r^{2},
\end{equation} 
where the functions $\mathcal{P}$, $\mathcal{Q}$ depend on $r$, $\mathcal{P} \geq 0$, $\mathcal{Q} \geq 0$. We choose the basis of null vectors\footnote{A similar null basis in three-dimensional spacetimes is used in Ref.\  \cite{LopezOrtega:2004cq}.} 
\begin{equation} \label{baseo}
\widehat{e}^{\ \mu}_{1} = \frac{1}{\sqrt{2}}\left( \frac{1}{\mathcal{P}},\frac{1}{\mathcal{Q}}\right) ,
 \ \ \ \ \
\widehat{e}^{\ \mu}_{2} = \frac{1}{\sqrt{2}}\left( \frac{1}{\mathcal{P}},-\frac{1}{\mathcal{Q}}\right), 
\end{equation}
that satisfies
\begin{equation}
 \widehat{e}_{a}^{\ \mu}\widehat{e}_{b \mu} = \eta_{ab} 
\end{equation} 
where $a,b = 1,2,$  and 
\begin{equation}
\left(\eta_{ab}\right) =\left(\eta^{ab}\right) =
\begin{pmatrix}
0 & 1 \\
1 & 0 \\
\end{pmatrix} .
\end{equation}

Using the representation of the gamma matrices
\begin{equation} \label{matgamma}
\gamma_{1} = \frac{1}{\sqrt{2}} \begin{pmatrix}
0 & 2 \\
0 & 0 \\
\end{pmatrix}, 
\,  \, \qquad
\gamma_{2} = \frac{1}{\sqrt{2}} \begin{pmatrix}
0 & 0 \\
2 & 0 \\
\end{pmatrix}, 
\end{equation}
that fulfills $\gamma_{a} \gamma_{b} + \gamma_{b}\gamma_{a} = 2 \eta_{ab} I$, we find that in the 2D spacetime (\ref{e: metric bidi PQ}) the Dirac equation simplifies to the coupled system of partial differential equations 
\begin{equation} \label{e: coupled partial Dirac}
\begin{split}
\frac{1}{\mathcal{P}} \partial_{t}\psi_{2} - \frac{1}{\mathcal{Q}}\partial_{r}\psi_{2}- \frac{1}{2\mathcal{P}\mathcal{Q}}\frac{d \mathcal{P}}{d r}\psi_{2} = -im\psi_{1} ,
\\
\frac{1}{\mathcal{P}} \partial_{t}\psi_{1} + \frac{1}{\mathcal{Q}}\partial_{r}\psi_{1} + \frac{1}{2\mathcal{P}\mathcal{Q}}\frac{d \mathcal{P}}{d r}\psi_{1} = -im\psi_{2},
\end{split}
\end{equation}
where $ \psi_{1}$ and $\psi_{2}$ are the components of the 2-spinor
\begin{equation}
\Psi = \begin{pmatrix}
\psi_{1}  \\
\psi_{2}  \\
\end{pmatrix}.
\end{equation}

Taking the components $ \psi_{1}$ and $\psi_{2}$ in the form 
\begin{equation}
\psi_{1} = R_{1}(r) e^{-i\omega t}, \qquad  \qquad \psi_{2} = R_{2}(r) e^{-i\omega t} ,
\end{equation}
from the coupled partial differential equations (\ref{e: coupled partial Dirac}) we get that the functions $R_1$ and $R_2$ must be solutions for the coupled system of first order differential equations 
\begin{align} \label{e: coupled radial odes}
 \frac{1}{\mathcal{Q}} \frac{d R_2}{d r}   + \frac{i \omega}{\mathcal{P}} R_2 + \frac{1}{2 \mathcal{P} \mathcal{Q}} \frac{d \mathcal{P}}{dr} R_2 &= i m R_1, \nonumber \\
  \frac{1}{\mathcal{Q}} \frac{d R_1}{d r} - \frac{i \omega}{\mathcal{P}} R_1  + \frac{1}{2 \mathcal{P} \mathcal{Q}} \frac{d \mathcal{P}}{dr} R_1 &= - i m R_2.
\end{align}

For 2D spacetimes that fulfill $\left(\mathcal{P}\mathcal{Q}\right)^2 = 1$, as the 2D black hole (\ref{e: black hole two-dimensional}), from the previous system of coupled differential equations for the functions $R_{l}$  we obtain the decoupled equations 
\begin{align} \label{e: eqs decoupled Dirac}
& \mathcal{P}^{4} \frac{d ^{2}R_{l}}{d r^{2}} + 2\mathcal{P}^{3}\frac{d \mathcal{P}}{d r}\frac{d R_{l}}{d r} \nonumber \\ 
&+ \left( \omega^{2} \pm i \omega \mathcal{P} \frac{d \mathcal{P}}{d r} + \frac{\mathcal{P}^{3}}{2}\frac{d ^{2} \mathcal{P}}{d r^{2}} + \left( \frac{\mathcal{P}}{2} \frac{d \mathcal{P}}{d r}\right)^{2} -\mathcal{P}^{2}m^{2}\right)R_{l} = 0.
\end{align}
We also point out that 
\begin{equation} \label{e: tortuga P}
 \frac{d ^{2} R_{l}}{d r_{*}^{2}} = \mathcal{P}^2\frac{d}{d r}\left(\mathcal{P}^{2}\frac{d R_{l}}{dr}\right)= \mathcal{P}^{4} \frac{d^{2}R_{l}}{dr^{2}} + 2 \mathcal{P}^{3}\frac{d \mathcal{P}}{d r}\ \frac{d R_{l}}{d r},
\end{equation}
and therefore Eqs.\  (\ref{e: eqs decoupled Dirac}) simplify to Schr\"odinger type equations
\begin{equation} \label{EqSchoDirac}
\frac{d^{2} R_{l}}{d r_{*}^{2}} + \omega^2 R_{l} = V_{l}R_{l},
\end{equation}
with effective potentials 
\begin{equation} \label{e: Potentials Dirac}
V_{l} = m^2\mathcal{P}^{2} \mp \frac{i \omega}{2}\frac{d \mathcal{P}^{2}}{d r} - \frac{\mathcal{P}^{2}}{4}\frac{d ^{2}\mathcal{P}^{2}}{d r^{2}} + \frac{1}{16}\left(\frac{d \mathcal{P}^{2}}{d r}\right)^{2} .
\end{equation}
In contrast to the effective potentials of Ref.\  \cite{LopezOrtega:2011sc} that involve square roots (see the formulas (\ref{e: potentials Dirac square root})), the previous effective potentials does not involve square roots of the metric functions. Furthermore, we notice that for the 2D black hole (\ref{e: black hole two-dimensional}) it is true that $\mathcal{P}^{2}=f = a^2 r^2 - 1/ (a r)$.

We also notice that taking 
\begin{equation}
 R_1=\frac{1}{P^{1/2}}\hat{R}_1, \qquad \qquad \qquad  R_2=\frac{1}{P^{1/2}}\hat{R}_2,
\end{equation} 
from the coupled system of differential equations (\ref{e: coupled radial odes}) we find that the functions $\hat{R}_1$ and $\hat{R}_2$ must be solutions of the coupled equations 
\begin{align} \label{e: coupled Dirac system hat}
 \frac{d\hat{R}_2 }{d r_*} + i \omega \hat{R}_2 = m P \hat{R}_1 , \qquad \qquad \qquad  
 \frac{d\hat{R}_1 }{d r_*} - i \omega \hat{R}_1 = m P \hat{R}_2,
\end{align}
where $r_*$ now denotes the tortoise coordinate of the metric (\ref{e: metric bidi PQ}) and is given by
\begin{equation}
 \frac{d  r_*}{d r} = \frac{Q}{P}.
\end{equation} 

Defining 
\begin{equation}
 Z_\pm = \hat{R}_1 \pm \hat{R}_2, \qquad \qquad \qquad W = m P,
\end{equation} 
from (\ref{e: coupled Dirac system hat}) we get that the functions $ Z_\pm$ satisfy 
\begin{align}
 \frac{d  Z_+}{d r_*} - W Z_+ = i \omega Z_-,  \qquad \qquad \qquad \frac{d  Z_-}{d r_*} + W Z_- = i \omega Z_+.
\end{align}
As is well known \cite{Chandrasekhar book}--\cite{Lopez-Ortega:2012ypj}, from the previous equations we find that the functions $Z_\pm$ are solutions of the Schrodinger type equations 
\begin{equation}
 \frac{d^2  Z_\pm}{d r_*^ 2} + \omega^ 2 Z_\pm - V_\pm Z_\pm = 0,
\end{equation} 
where the effective potentials $V_\pm$ are equal to
\begin{equation} \label{e: superpotential potentials}
 V_\pm = W^2 \pm  \frac{d  W}{d r_*} .
\end{equation} 
Explicitly, in the spacetime (\ref{e: metric bidi PQ}) they are equal to
\begin{equation}
 V_\pm = m^2 P^2 \pm m \frac{d  P}{d r_*}.
\end{equation} 
From the expressions (\ref{e: superpotential potentials}) we see that the effective potentials $V_\pm$ are supersymmetric partners \cite{Cooper:1994eh}, \cite{Lopez-Ortega:2012ypj} (see Ref.\ \cite{LopezOrtega:2011sc} for a similar result in static diagonal 2D black holes, but in a different dyad of basis vectors). As a consequence, we expect that the functions $R_1$ and $R_2$ have the same spectrum. It is convenient to notice that we have not been able to write the effective potentials (\ref{e: Potentials Dirac})  in the form (\ref{e: superpotential potentials}), that is, we have not found a function $W$ that is a superpotential for the effective potentials (\ref{e: Potentials Dirac}). Despite this fact, we think that the previous arguments indicate that the two potentials  (\ref{e: Potentials Dirac})  have the same spectrum (see also Sect.\ \ref{ss: numerical Dirac} and Table \ref{Tabla-V1+V2}). 

Finally, we note that in a similar way to the previous results for 2D static black holes, it is known that in higher dimensional static black holes, the Dirac equation simplifies to a pair of Schr\"odinger type equations, see Refs.\ \cite{Chandrasekhar book}--\cite{Destounis:2018qnb} for some examples.

\section{Quasinormal frequencies of the Klein-Gordon field from a coupled system}
\label{a: KG coupled AIM}

In Sect.\ \ref{ss: aim Klein-Gordon} we use the formulation of the AIM for second order differential equations \cite{Ciftci Hall and Saad 2003} to calculate the QNFs of the Klein-Gordon field. Taking into account that a second order linear differential equation can be written as a coupled system of first order differential equations \cite{Boyce-book}, we can exploit the AIM for coupled systems of first order differential equations to compute its QNFs. 

To show this fact we rewrite Eq.\ (\ref{e: radial AIM Klein-Gordon}) as
\begin{equation} \label{e: radial Klein-Gordon coupled}
 \frac{d^2 \hat{R} }{d z^2} = \lambda_0 \frac{d \hat{R} }{d z} + s_0 \hat{R},
\end{equation} 
where $\lambda_0$ and $s_0$ are given in the expressions (\ref{e: lambda 0 ese 0}). Defining the functions $W_1$, $W_2$ by
\begin{equation}
 W_1  =  \hat{R}, \qquad \qquad W_2 = \frac{d \hat{R} }{ d z},
\end{equation} 
that is, $d^2 \hat{R} /dz^2=dW_2/dz$, we encounter that Eq.\ (\ref{e: radial Klein-Gordon coupled}) is equivalent to the coupled  system of first order differential equations 
\begin{align}
 \frac{dW_1}{dz} &= W_2 , \\
 \frac{dW_2}{dz} &=  s_0 W_1  + \lambda_0 W_2 . \nonumber
\end{align}
Comparing to the system of first order differential equations (\ref{e: aim coupled}) of Sect.\ \ref{ss: aim Dirac}, for the Klein-Gordon field we find that the functions $\Lambda_0$, $ S_0$, $\Omega_0$, $P_0$ are equal to
\begin{align}
 \Lambda_0 =0  , \qquad \qquad 
 S_0 = 1 , \qquad  \qquad 
 \Omega_0  = s_0  , \qquad \qquad 
 P_0 = \lambda_0.
\end{align}

Taking as a basis these functions, we use the recurrence relations (\ref{e: recurrence relations coupled}) to calculate the quantities that appear in the quantization condition (\ref{e: delta coupled}), whose stable roots are the QNFs of the Klein-Gordon field. Implementing for the Klein-Gordon field the  AIM for coupled systems of first order differential equations as previously commented, we find that this approach to the problem yields the same values for its QNFs that we present in Sect.\ \ref{ss: numerical Klein-Gordon}. Therefore we think that this transformation can be used for any second order differential equation of the form (\ref{e: radial Klein-Gordon coupled}) and we can calculate its eigenvalues with the AIM for coupled systems of first order differential equations. Moreover, we think that this alternative formulation can be used to verify the results that produces the AIM for second order differential equations.

\section{The improved asymptotic iteration method}
\label{a: iaim coupled}

In the AIM for second order differential equations, the recurrence relations (\ref{eq: recurrence general}) demand the calculation of the derivatives for the functions $\lambda_n$, $s_n$, a process that requires many resources. To make the AIM more efficient in the computation of QNFs, in Ref.\ \cite{Cho:2009cj} (see also \cite{Cho:2011sf}) it is proposed the improved asymptotic iteration method (IAIM in what follows) that allows us to determine the QNFs without computing the derivatives of the functions $\lambda_n$, $s_n$. In the IAIM we first expand the functions $\lambda_n$, $s_n$ around a point $\xi$ in the form \cite{Cho:2009cj}, \cite{Cho:2011sf}
\begin{equation}\label{e: serie lambda iaim}
\lambda_n (\xi )= \sum ^\infty_{i=0} c^i_n (x- \xi )^i, \qquad 
s_n (\xi )= \sum^\infty_{i=0} d^i_n (x- \xi )^i,
\end{equation}
and then we substitute these expressions into the recurrence relations (\ref{eq: recurrence general}) to find that the coefficients  $c^i_n$,  $d^i_n$ satisfy \cite{Cho:2009cj}, \cite{Cho:2011sf}
\begin{equation}\label{e:coef c d iaim}
c^i_n = (i+1)c^{i+1}_{n-1} + d^i_{n-1} + \sum^i_{k=0} c^k_0 c^{i-k}_{n-1}, \qquad 
d^i_n = (i+1) d^{i+1}_{n-1} + \sum ^i_{k=0} d^k_0 c^{i-k}_{n-1}.
\end{equation}
Using these coefficients, we write the quantization condition (\ref{eq: quantization})  as follows \cite{Cho:2009cj}, \cite{Cho:2011sf}
\begin{equation}\label{e: final IAIM}
d^0_n c^0_{n-1} - d^0_{n-1} c^0_n =0.
\end{equation}
In a similar way to the AIM, the stable roots of this equation are the QNFs of the field. Taking as a basis Eq.\ (\ref{e: radial AIM Klein-Gordon}),  if we use the IAIM to calculate the QNFs of the Klein-Gordon field in the 2D black hole (\ref{e: black hole two-dimensional}), then we obtain the same numerical results for the QNFs that we expound in Sect.\ \ref{ss: numerical Klein-Gordon}.

Based on this extension of the AIM for second order differential equations, in what follows we  develop an improved version of the AIM for coupled systems of first order differential equations, since in this method  we also require the calculation of the derivatives  for the functions $\Lambda_n$, $S_n$, $\Omega_n$, $P_n$ to evaluate the recurrence relations (\ref{e: recurrence relations coupled}) and as previously commented  these mathematical operations demand many resources. As far as we know, this extension of the AIM for coupled systems has not previously presented. We first expand the functions  $\Lambda_0$, $S_0$, $\Omega_0$, and $P_0$ around a point $\xi$ (in a similar way to the formulas (\ref{e: serie lambda iaim}) for the AIM of second order differential equations)
\begin{align}
\Lambda_n(\xi) =  \sum_{i=0}^\infty C_{n}^{i} (x-\xi)^i , \qquad \Omega_n(\xi)=\sum_{i=0}^\infty E_{n}^{i} (x-\xi)^i , \nonumber \\ 
S_n(\xi)=\sum_{i=0}^\infty D_{n}^{i} (x-\xi)^i , \qquad P_n(\xi)=\sum_{i=0}^\infty F_{n}^{i} (x-\xi)^i, 
\end{align} 
and then we substitute these expressions in the recurrence relations (\ref{e: recurrence relations coupled}), to get  that the coefficients $C_{n}^{i}$, $D_{n}^{i} $, $E_{n}^{i}$, $F_{n}^{i}$, satisfy the recurrence relations\footnote{We point out that the mathematical structure of the recurrence relations (\ref{e: recurrence couple iaim}) is more symmetric than that of the expressions (\ref{e:coef c d iaim}). } 
\begin{align} \label{e: recurrence couple iaim}
 C_{n}^{i} &= \sum_{k=0}^i \left( C_{n-1}^{i-k} C_{0}^{k} + D_{n-1}^{i-k} E_{0}^{k}  \right) + (i+1) C_{n-1}^{i+1} , \nonumber \\ 
 D_{n}^{i} &= \sum_{k=0}^i \left( C_{n-1}^{i-k} D_{0}^{k} + D_{n-1}^{i-k} F_{0}^{k}  \right) + (i+1) D_{n-1}^{i+1} , \nonumber \\ 
 E_{n}^{i} &= \sum_{k=0}^i \left( E_{n-1}^{i-k} C_{0}^{k} + F_{n-1}^{i-k} E_{0}^{k}  \right) + (i+1) E_{n-1}^{i+1} , \nonumber \\ 
 F_{n}^{i} &= \sum_{k=0}^i \left( E_{n-1}^{i-k} D_{0}^{k} + F_{n-1}^{i-k} F_{0}^{k}  \right) + (i+1) F_{n-1}^{i+1} .
\end{align}

Using these coefficients, we get that the quantization condition (\ref{e: delta coupled}) of the AIM for systems of coupled first order differential equations takes the form
\begin{equation}
 D_{n}^{0} C_{n-1}^{0} -  D_{n-1}^{0} C_{n}^{0} = 0.
\end{equation} 
This condition is similar to Eq.\ (\ref{e: final IAIM}) of the AIM, but we notice that the coefficients which appear in the last condition satisfy different recurrence relations than the coefficients $c^i_n$, $d^i_n$. Taking as a basis the expressions (\ref{e: lambda s omega p coupled}), if we use this improved formulation of the AIM for coupled systems of first order differential equations to compute the QNFs of the Dirac field propagating in the 2D black hole (\ref{e: black hole two-dimensional}), then we obtain the same QNFs given in Sect.\ \ref{ss: numerical Dirac}.

\section{Quasinormal frequencies of the massive Dirac field in a $D$-dimensional Lifshitz black hole\footnote{We thank one of the reviewers for suggesting that we analyze whether the coupled AIM works to calculate the QNFs of the Dirac field moving in the $D$-dimensional Lifshitz black hole studied in Refs.\ \cite{Balasubramanian:2009rx}--\cite{Lopez-Ortega:2014daa}. }}
\label{a: Lifshitz qnfs}

The dynamics of classical fields has been studied in the $D$-dimensional Lifshitz black hole with metric \cite{Balasubramanian:2009rx}-\cite{Lopez-Ortega:2014daa}
 \begin{equation} \label{e: black hole Lifshitz}
d s^2 = \frac{r^4}{l^4} \left( 1- \frac{r_+^2}{r^2} \right) d t^2 - \frac{l^2 d r^2}{r^2 -r_+^2} - r^2  d \Omega^2 ,
\end{equation} 
where $r_+$ is the radius of the event horizon, $d  \Omega^2$ is the metric of a $(D-2)$-dimensional plane, and $l$ is a  constant that determines the Lifshitz radius \cite{Balasubramanian:2009rx}, \cite{Giacomini:2012hg}. In this Lifshitz black hole the exact values of the QNFs for several fields are known \cite{Giacomini:2012hg}--\cite{Lopez-Ortega:2014daa}. In particular, in the zero momentum limit, the QNFs of the massive Dirac field with mass $m$ are  exactly calculated and they are given by \cite{Catalan:2013eza}, \cite{Lopez-Ortega:2014daa}
\begin{equation} \label{e: QNF Dirac Lifshitz}
 \tilde{\omega} = - i \left( n + \frac{\tilde{m}}{2} +  \frac{1}{2} \right),
\end{equation}
where  $\tilde{m} = m l$,  $\tilde{\omega} = (\omega l^3)/r_+^2$, and $n=0,1,2,\ldots$

To investigate whether the coupled AIM works to determine the QNFs of the massive Dirac field moving in the $D$-dimensional Lifshitz black hole (\ref{e: black hole Lifshitz}), in what follows we show how to transform the coupled radial equations that govern its dynamics. As shown in Ref.\ \cite{Lopez-Ortega:2014daa}, in the  $D$-dimensional Lifshitz black hole (\ref{e: black hole Lifshitz}) the massive Dirac equation in the zero momentum limit takes the form 
\begin{align} \label{e: Dirac radial kappa}
 z(z^2 -1) \frac{d R_2}{d z} + i \tilde{\omega} R_2 &=  z (z^2 -1)^{1/2} i \tilde{m} R_1 ,   \\ 
z(z^2 -1) \frac{d R_1}{d z} - i \tilde{\omega} R_1 &= - z (z^2 -1)^{1/2} i \tilde{m} R_2,\nonumber 
\end{align}
where $z= r/r_+$ with $z \in (1,\infty)$. 

To transform Eqs.\ (\ref{e: Dirac radial kappa}) into a convenient form we define the functions $W_1$ and $W_2$ by
\begin{equation}
 R_1=\frac{(z^2-1)^{1/2}}{z}W_1, \qquad \qquad R_2=W_2,
\end{equation} 
to find that they must be solutions of the coupled system of first order differential equations
\begin{align}
 & z(z^2 -1) \frac{d W_2}{d z} + i \tilde{\omega} W_2 =   (z^2 -1) i \tilde{m} W_1 ,   \\
& z(z^2 -1) \frac{d W_1}{d z} +(1- i \tilde{\omega}) W_1 = - z^2 i \tilde{m} R_2.\nonumber
 \end{align} 
To consider the boundary conditions of the QNMs we take the functions $W_1$ and $W_2$ as
\begin{equation}
 W_1=h(z) U_1, \qquad \qquad  W_2 = h(z) U_2 ,
\end{equation} 
with 
\begin{equation}
 h(z)=(z-1)^{-i \tilde{\omega}/2} z^{-\tilde{m}+i \tilde{\omega}/2},
\end{equation} 
to find that the functions $U_1$ and $U_2$ satisfy the system of differential equations 
\begin{align}
 \frac{d U_1}{d u} &= \left( \frac{i \tilde{\omega} }{2u} + \frac{ \tilde{m} }{1-u} +\frac{(i \tilde{\omega} -1)(1-u)}{u(2-u)} \right) U_1 - \frac{i \tilde{m} }{u(2-u)(1-u)} U_2 ,\\
 \frac{d U_2}{d u} &= \frac{i \tilde{m} }{(1-u)} U_1 + \left( \frac{i \tilde{\omega}}{2u} + \frac{ \tilde{m} }{1-u} - \frac{i \tilde{\omega}(1-u)}{u(2-u)} \right) U_2, \nonumber
\end{align}
where the variable $u$ is defined by
\begin{equation}
 u=\frac{z-1}{z} .
\end{equation} 

This system of equations is of the form (\ref{e: aim coupled}) and as a consequence, we can identify the functions $\Lambda_0$, $S_0$, $\Omega_0$, and $P_0$ as follows
\begin{align}
 \Lambda_0 &= \frac{i \tilde{\omega} }{2u} + \frac{ \tilde{m} }{1-u} +\frac{(i \tilde{\omega} -1)(1-u)}{u(2-u)} ,\nonumber \\  
S_0 &= - \frac{i \tilde{m} }{u(2-u)(1-u)} ,\nonumber \\  
\Omega_0 &=  \frac{i \tilde{m} }{(1-u)},\\   
P_0 &=  \frac{i \tilde{\omega}}{2u} + \frac{ \tilde{m} }{1-u} - \frac{i \tilde{\omega}(1-u)}{u(2-u)} .\nonumber
\end{align}
From these functions we can use the coupled AIM to determine, in the zero momentum limit, the QNFs of the massive Dirac field propagating in the $D$-dimensional Lifshitz black hole (\ref{e: black hole Lifshitz}). In Table \ref{Tabla-Lifshitz} we compare the QNFs that we obtain from the coupled AIM and those that we get from the analytical expression (\ref{e: QNF Dirac Lifshitz}). We see the QNFs of the first ten QNMs coincide very well.

\begin{table}[ht]
\centering
\caption{ For  the $D$-dimensional Lifshitz black hole (\ref{e: black hole Lifshitz}) we show the QNFs of the Dirac field that produce the analytical expression (\ref{e: QNF Dirac Lifshitz}) and the coupled AIM. We take $\tilde{m} = 3/7$.}
\begin{tabular}{|c|c|c|}
\hline
Mode number & Analytical & Coupled AIM \\ \hline
 0 & $-0.7143 i$ &  $-0.7143 i$ \\ \hline
 1 & $-1.7143 i$ &  $-1.7143 i$ \\ \hline
 2 & $-2.7143 i$ &  $-2.7143 i$ \\ \hline
 3 & $-3.7143 i$ & $-3.7143 i$ \\ \hline
 4 & $-4.7143 i$ &  $-4.7143 i$ \\ \hline
 5 & $-5.7143 i$ &  $-5.7143 i$ \\ \hline
 6 & $-6.7143 i$ &  $-6.7143 i$ \\ \hline
 7 & $-7.7143 i$ &  $-7.7143 i$ \\ \hline
 8 & $-8.7143 i$ &  $-8.7143 i$ \\ \hline
 9 & $-9.7143 i$ &  $-9.7143 i$ \\ \hline
\end{tabular} 
\label{Tabla-Lifshitz}
\end{table}

\end{document}